# Reference Capabilities for Safe Parallel Array Programming

Beatrice Åkerblom[a], Elias Castegren[b], and Tobias Wrigstad[c]

a   Stockholm University
b   KTH Royal Institute of Technology
c   Uppsala University

**Abstract**   The array is a fundamental data structure that provides an efficient way to store and retrieve non-sparse data contiguous in memory. Arrays are important for the performance of many memory-intensive applications due to the design of modern memory hierarchies: contiguous storage facilitates spatial locality and predictive access patterns which enables prefetching.

Operations on large arrays often lend themselves well to parallelisation, such as a fork-join style divide-and-conquer algorithm for sorting. For parallel operations on arrays to be deterministic, data-race freedom must be guaranteed. For operations on arrays of primitive data, data-race freedom is obtained by coordinating accesses so that no two threads operate on the same array indices. This is however not enough for arrays of non-primitives due to aliasing: accesses of separate array elements may return pointers to the same object, or overlapping structures.

Reference capabilities have been used successfully in the past to statically guarantee the absence of data-races in object-oriented programs by combining concepts such as uniqueness, read-only access, immutability, and borrowing. This paper presents the first extension of reference capabilities—called array capabilities—that support concurrent and parallel operations on arrays of both primitive and non-primitive values.

We define a small language of operators on array capabilities. In addition to element access, these operations support the abstract manipulation of arrays: logical splitting of arrays into subarrays; merging subarrays; and in-place alignment of the physical elements of an array with the logical order defined through splitting and merging. The interplay of these operations with uniqueness, borrowing and read-only access allows expressing a wide range of array use cases.

By formalising our type system and the dynamic semantics of the key operations on array capabilities in a simple calculus, we give a precise description of the array capabilities. We when prove type soundness and array disjointness—that two capabilities that allow mutation can never give access to overlapping sets of elements. Data-race freedom is stated as a corollary to these theorems, as data races require two array capabilities in different threads. In addition to the formal approach, we experiment with array capabilities on a small number of parallel array algorithms and report our preliminary findings.

Array capabilities extend the safety of reference capabilities to arrays in a way that build cleanly on existing reference capability concepts and constructs. This allows programmers to express parallel array algorithms on a higher level than individual indices, and with absences of data races guaranteed statically.

## The Art, Science, and Engineering of Programming

**Reference Capabilities for Safe Parallel Array Programming**

# 1 Introduction

The array is a fundamental data structure, used across countless programs since the infancy of programming. In addition to providing an efficient way to store and retrieve non-sparse data by way of indexing, arrays provide a means of storing data contiguously in memory. This is important for performance due to how modern cache systems are constructed, as well as the increasing gap between CPU performance and memory bandwidth [28], exacerbated by an increasing number of cores.

Operations on large arrays of data are typically carried out using multiple cores operating on disjoint ranges of indexes. Parallel array operations are relatively straightforward—the problem size is known ahead of time, and classical divide-and-conquer algorithms can logically subdivide arrays into smaller chunks until they fit in cache. Operations on multidimensional arrays may cause less intuitive access ranges (e.g., row major vs. column major orientation and access patterns), or adjusting indexes in ranges to avoid false sharing problems where accesses to disjoint array cells located on a common cache line cause performance degradation due to coherency traffic. In many cases, typical off-by-one errors may cause data races, which—like most concurrency bugs—can be hard to reproduce and debug.

Data-races require aliasing: without the ability to give multiple names for the location of a datum, two or more threads cannot be racing on it. Because of the efficiency of sharing of data by pointer rather than by copy, researchers in the object-oriented setting have developed reference capability systems to mitigate aliasing in concurrent programs. Some such systems only prevent aliasing of mutable state [6, 13, 27] while others resort to dynamic means, e.g., forcing the use of locks or transactions for accesses to shared data [4, 11, 22] or even lock-free protocols [12]. Reference capabilities are in this respect different from more traditional effect systems [23] which freely permit all aliasing, but control their concurrent usage (e.g., [3, 16]).

Sharing arrays safely across multiple threads require that read–write or write–write accesses do not overlap, meaning that *indexes for such access must not alias*. This effectively reintroduces the problems of pointer arithmetic in reference-based systems and is a source of difficulty for effect-based concurrency control: effect disjointness turns into static reasoning about whether expressions may evaluate to the same integers. This is visible for example in Deterministic Parallel Java [2] which resorts to dynamic checking of indexes to prove effect disjointness.

In this paper, we explore a reference-capability-based approach to safe array handling in concurrent applications. We start from *Kappa* [9], which supports a rich taxonomy of reference capabilities, and extend it with the aim of enabling data race-free parallel programming with support for arrays. The resulting system, *Arr-O-Matic*, supports array operations which are rich enough to express typical parallel operations on arrays, such as parallel sorting algorithms, stencil operations, and blocked matrix operations. Through integration with a standard type system, *Arr-O-Matic* provides a compile-time guarantee of data-race freedom. In particular, we:

- introduce array capabilities (§ 3), an extension of reference capabilities, that provides built-in support for subdivision of arrays into disjoint parts to enable data

- race-free parallel programming with support for arrays. We do this in the context of *Arr-O-Matic*, an extension of Arrgh [1];
- examine how array capabilities can support abstract manipulation of arrays through logical splitting into subarrays and merging of subarrays (§ 3.4–3.6);
- apply our array capabilities and associated operations on several examples that exercise realistic properties of parallel array algorithms (§ 3.7);
- present selected parts of the formal description of *Arr-O-Matic* with a focus on arrays and concurrency (§ 4);
- give the static and dynamic semantics (§ 4.1–4.2);
- state and prove the important meta-theoretic properties including our key invariant, "disjointness of arrays" (§ 4.3),
- briefly state how we have tested our design on existing problems (§ 5).

Section 2 introduces the necessary building blocks: arrays, typical array programs, matrices, and reference capabilities. Section 6 covers related work. Section 7 concludes.

## 2 Background: Programming with Arrays (and Parallelism)

In this section, we overview common array usages and language support for arrays (more specific comparisons can be found in § 6), and in particular array slices. We specifically discuss parallelism when programming with arrays, and the shortcomings of using reference capabilities, a technique for excluding data-race bugs, with arrays.

**Typical Array Operations**  The typical array stores zero or more values of a single type consecutive in memory. The canonical operations on arrays are reading and writing individual elements by index. Many programming languages index arrays from zero, some from one. A common programming mistake is an off-by-one error [14], typically iterating one time too many or too few in a loop accessing elements in an array by index. Such an error is commonly related to using "less than or equal to" in place of a "strictly less than", or similar. Depending on the nature of the language (and whether it is off by $-1$ or $+1$), such bugs can be tricky to find.

Many modern languages allow foreach-style iteration over arrays without the need for explicit element accesses or indexes (making off-by-one errors less likely). For example, Java allows iterating over an array of strings thus: **for** (String s : myStrings) ... Because foreach loops do not directly manipulate indexes (cf. ++i), they need not give an ordering guarantee. This simplifies implicit parallel execution of loops, for example in Fortress [26]. Such parallel operations on arrays of primitive data are trivially safe because all array elements contain different (copies of) values. However, operating on arrays of references to complex objects in parallel requires care to make sure that references are not aliases and that the objects do not share mutable state.

**Divide-and-Conquer Algorithms**  A common strategy for (parallel) array algorithms is divide-and-conquer, used for example in quicksort as shown in listing 1.

**Reference Capabilities for Safe Parallel Array Programming**

■ **Listing 1** Quicksort implementations. Right: Python. Left: Encore, an imperative, object-oriented actor programming language. For familiarity, we have changed the Encore syntax from **do**, **then** and **end** to brackets. **async** spawns a new task.

```
1  def qs( int[]  arr, int lo, int hi ) {
2     if (hi > lo) {
3        int pivot = lo
4        int pivPos = partition( arr, lo, hi, pivot )
5        async { qs( arr, lo, pivPos ) }
6        qs( arr, pivPos + 1, hi )
7     }
8  }
```

```
1  def qs(arr):
2     if len(arr) > 1:
3        pivot = arr[0]
4        pivPos = partition(arr,  pivot)
5        return qs(arr[:pivPos-1]) + \
6           [pivot] + qs(arr[pivPos+1:])
7     else:
8        return arr
```

In listing 1, each recursive step calls the qs() function on smaller portions of the array, spawning a new task for one portion and reusing the current task for the other. Consequently, possibly parallel tasks will have concurrent write access to the array. This in turn means that the soundness of the implementation depends on the programmer correctly using index manipulations to control what parts of the array will be accessed by each thread. In the implementation in listing 1, the code keeps track of its current portion of the array through start and end index values. It subsequently uses the index of the pivot to calculate the two new array portions for the recursive calls. All in all, six different indexes are involved. Off-by-one errors on any of these will not be detected as out of bounds accesses of the logical array portion, but may be subject to data-races which can lead to the array not being properly sorted.

To decrease the reliance on indexes, some languages (originally Fortran, but later e.g., APL, Rust and Julia) support *array slicing*. Slicing is an operation that can be used on data structures like arrays or lists to obtain a substructure of the original. Given a Python array, [2, 4, 6, 8], the slicing operation [1:3] returns the array with only the elements in the range $(1, 3)$, i.e., [4, 6]. In the case of Python, the slicing operation returns a new array with copies of the elements, but other languages like D and Julia support slicing that returns a cropped alias, a logical, zoomed in "view" of the same array. Operations on this logical array view are reflected on the underlying physical array. Rust also supports slicing, but to relax its restrictions on mutating aliased data, this is managed through a library that wraps unsafe code whose correctness is manually verified, but whose use is guaranteed to be correct.

Array slices allow manipulating a subpart of an array without keeping track of start and stop indexes. The rightmost code in listing 1 shows a (sequential) implementation of quicksort using array slices in Python. The concatenation on lines 5 and 6 and the return on line 8 are required because Python's array slices are created by copy.

**Arrays as Matrices**   Matrices implemented as multidimensional arrays are common in scientific computing. For example, HAParaNDA [21], an iterative solver for high-dimensional linear partial differential equations implemented in thousands of lines of C++ spends almost all execution time on stencil applications on matrices.

Because of the stencil shape, a stencil application on a matrix cannot follow the divide-and-conquer approach, as writes typically depend on reads on adjacent ele-

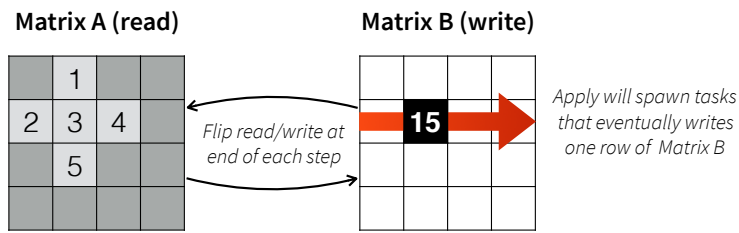

■ **Figure 1** Stencil application from read-only matrix to write-only matrix in HAParaNDA. Once all cells in Matrix *B* are written, the read and write roles are flipped.

■ **Listing 2** Part of the central stencil code of HAParaNDA, translated into Encore.

```
def apply( from : Matrix[int],  to : Matrix[int],  start : int, stop : int ) : unit {
  if ( stop != start ) {
    async { apply( from, to, start, ( stop - start ) / 2 ) }
    apply( from, to, start + (stop - start) / 2, stop )
  } else {
    do_row( from, to, start )
  }
}

// Apply stencil  to  one row
def do_row( from : Matrix[int], to : Matrix[int],  y : int ) : int {
  for x <- [0..from.length()]  { to[x][y]  = read_stencil( from, x, y ) }
}

// read from the from matrix, wrapping around for out of bounds indexes
def read_stencil( from : Matrix[int],  x : int,  y : int ) : int { … }
```

ments. To this end, HAParaNDA uses two matrices, one read-only matrix and one write-only matrix, whose roles are flipped to run in constant space. Figure 1 shows this pictorially, using a plus-shaped stencil and storing the sum of the elements of the stencils in the write-only array. Listing 2 shows a simplified version of the parallel stencil application translated into Encore code from C++.

The read-only matrix is safe to share across all parallel threads, but, as with the quicksort example, great care must be taken to ensure that no two threads may write to the same place in the write-only matrix. Furthermore, care must be taken not to flip the roles of the two matrices until all writes have finished.

**Applying Reference Capabilities to Parallel Programming with Arrays**   As demonstrated by the parallel implementations of quicksort and stencil application, the common mistakes like off-by-one errors can introduce subtle data race bugs. Furthermore, mediating between read and write access modes for a datum requires guaranteeing the absence of outstanding operations, and updating all aliases' view of that datum. As demonstrated by the Python quicksort implementation, programming with array slices can simplify programming with arrays by reducing the number of explicitly manipulated indexes.

**Reference Capabilities for Safe Parallel Array Programming**

Reference capabilities have been used in the past to simplify concurrent programming [7, 9, 10, 11, 12, 13, 15] by enabling the programmer to express, in a statically tractable manner, how an object may be accessed through its incoming references, e.g., there is only a single reference to it which must be manipulated carefully to maintain its uniqueness, or there are only references through which reading but not writing is allowed, and rules for mediating between these situations.

A reference capability is essentially a normal reference with some restrictions on usage and how other references may be created from it. For example a *unique* reference capability is the only reference in a program to the object it refers, and a *read* reference capability is a reference that cannot be used to cause or witness mutation of the object it refers. Both these simple capabilities allow for some powerful reasoning and optimisations: accesses are oblivious to parallelism, caching of intermediate values or delaying synchronisation with global memory is safe, etc. Static propagation of capability metadata typically happens through types which allows a programmer or a compiler to locally reason about all accessible values. A well-typed program using reference capabilities is statically guaranteed to be data race-free.

Reference capabilities are expressive. They support e.g., divide-and-conquer-style parallel algorithms on complex object structures, by following left and right pointers of a binary tree that each give exclusive access right to their reachable sets of objects, or by decomposing a pair into disjoint first and second components. Through their combination with borrowing [5], a technique for temporarily relaxing constraints on manipulation of a capability, reference capabilities can be integrated in typical programming environments [8, 17, 25].

This work takes as its starting point the *Kappa* reference capability system [10]. *Kappa* has been implemented in the Encore programming language [8], which thus enjoys data-race freedom at compile-time, *with one caveat: arrays*. The initial *Kappa* work proposes to model arrays as objects with enumerable fields, but this does not allow divide-and-conquer-style decomposition. For this reason, Encore arrays are designated as *unsafe*—an escape hatch—and sharing arrays between threads results in compile-time warnings.

When applying *Kappa* reference capabilities to parallel array programs, it became clear that things were missing. A single array could be candidate for being decomposed in different ways under different circumstances (e.g., splitting a matrix into columns or rows), and a logic was missing for merging split parts. Furthermore, the capabilities did not model the consecutive nature of arrays, and so could not be used to reason about the difference between logical views and the underlying physical representation.

This paper extends *Kappa* with *array capabilities*, which support typical array programming patterns, and integrate well with the existing *Kappa* capabilities, to allow them to be used inside arrays.

## 3  Language Design

In this section we present the design of a set of coherent, orthogonal operations for concurrent and parallel programming with arrays based on a notion of *array* *capabilities* which extends the reference capability concept to handle aliasing of array indexes. Our proposal extends the *Kappa* reference capability system [9] and builds on our work on Arrgh [1]. We name our system *Arr-O-Matic*.

### 3.1 Array Capabilities

The core contribution of this work is the notion of array capabilities, which are similar to reference capabilities, but provide built-in support for subdivision into array slices. Like object capabilities, an array capability is an unforgeable token that governs access to a resource (an array). Like reference capabilities, array capabilities come with guarantees on what concurrent holders of overlapping array capabilities cannot do in terms of access to elements or contents stored in elements.

An array capability is an abstraction of an array. It may not give access to all elements of its underlying array, or even full access to the elements. Furthermore, in a departure from traditional arrays, adjacent indexes do not necessarily denote adjacent elements. All these aspects are transparent to a programmer and creating a new array returns an array capability that gives access to the newly created array. In the following, we will use array and array capability interchangeably, clearly pointing out when we mean the physical representation or logical representation (capability-level).

In addition to the typical array operations—accessing individual elements by index and taking the size—array capabilities support a coherent set of orthogonal operations through which a large number of all parallel array algorithms can be expressed. These operations are described in the next section.

### 3.2 Operations on Array Capabilities

We outline the specific operations on array capabilities below. §4 gives a formal account of a subset of their semantics.

**Duplicate** Creates a copy of an array capability: e.g., x = y creates a copy of the capability in y and stores it in x.

**Move** Like duplication, but nullifies the source: e.g., x = y atomically copies y's contents to x and then stores **null** in y. Movement preserves *uniqueness* (see below). Movement is equivalent to destructive reads [20].

**Split** Converts one array capability into several *sibling* array capabilities. Splitting maintains exclusive access by partitioning the underlying array capability into disjoint sub-arrays.

The partitioning of a split is logical, i.e., happens at the capability-level. This means that the underlying array is never affected by splitting and that capabilities merely act as proxies for the underlying array. Figure 2 shows this pictorially. Additionally, it shows how splitting can be *consecutive* or *strided,* to support typical array usage.

**Merge** Dual of splitting. Merge combines two or more array capabilities to form a single, new, array capability that governs access to the combined resources. Merge can use *concatenation* (dual of split's consecutive) or *interleaved* (dual of split's strided) semantics.

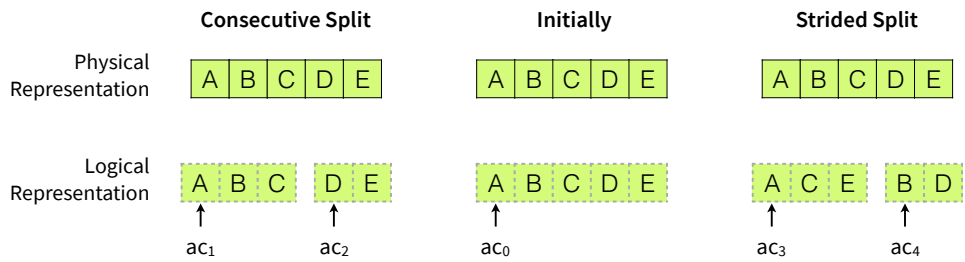

**Figure 2** Splitting array capabilities. The middle figure shows the initial state, e.g., after the creation of the array. Note that the physical representation and the logical representation ($ac_0$) are aligned. The left figure shows a consecutive 2-split—dividing the array into two consecutive subarrays ($ac_1$ and $ac_2$) of the original array. The right figure shows a strided 2-split, where the two subarrays ($ac_3$ and $ac_4$) are created from the odd and even elements of the array respectively.

Like split, merging is logical, meaning that the newly created capability simply acts as a proxy for the underlying array. (For simplicity, we leave merging capabilities accessing different underlying arrays for future work.)

**Align** Whereas logical partitioning suffices to guarantee data-race freedom, some algorithms enjoy a performance boost from physical partitioning e.g., to avoid false sharing, or to better align access patterns with caches. The align operation alters the physical representation of an array so that it aligns with a logical view, possibly creating a new array as a result.

**Borrow** Ties the life-time of a capability, and capabilities derived from it through splitting, to some lexical scope. This allows weakening and splitting of resources and stitching them back together in a structured programming fashion. For example, a typical pattern involves borrowing a resource, splitting it recursively, and finally exiting the scope of the borrowing which automatically reinstates the original capability without forcing the programmer to manually merge all siblings.

## 3.3 Access Modes

An array capability comes with an *access mode* that controls how data of the underlying resource is accessed and controls how the capability may be treated. These modes are also in *Kappa* [9], but are reinterpreted for arrays, where needed. Operations like splitting, borrowing, etc. preserve access modes.

**Unique** An array capability with unique access mode gives exclusive access to its underlying array and may therefore not be duplicated. Unique capabilities may not be duplicated, only moved.

**Read** An array capability with read access mode gives read-only access to its underlying array and is therefore safe to duplicate.

**Locked** An array capability with locked access mode requires accesses to array elements to be synchronised on a lock, which allows expressing algorithms that need to share mutable elements. Locks can be fine-grained (per element) or coarse-grained (per overlapping array capability) as desired by the application.

**Local** An array capability with local access mode denotes thread-exclusive access. Such a capability is safe to duplicate inside a thread but may not be copied to or moved to another thread (or whatever the unit of parallelism).

*All array capabilities guarantee the absence of data races in a well-typed program.* For an in-depth discussion of how to handle storing one type of capability inside another, see [9], which also includes a discussion of additional access modes. In short, the only way that two reference capabilities in different threads can be used to access the same data is if the data is immutable, or if the accesses are mediated by a lock, whose acquisition and release is statically enforced.

Having briefly overviewed the operations and access modes of array capabilities, the next section discusses the operations in more detail and finishes off with a set of examples that exercise them.

### 3.4 Maintaining Exclusivity through Splitting

Exclusive array capabilities are split by consuming the original array capability in return for *n* new array capabilities that govern access to disjoint parts of what the original capability could access.

Expressed as a function, the split operation takes the following form:

**split**( arr : Array[t], splits : **int**, strided : **bool** ) : Array[Array[t]]

The argument arr is an array capability for governing access to an array of t's and the argument splits denotes how many array capabilities arr should be split into. Finally, the argument strided controls whether to split the array in consecutive or strided subarrays. For example, the following splits an array of 5 elements into two subarrays of consecutive elements (leftmost example of Figure 2):

**split**( [A, B, C, D, E], 2, **false** ) → [ [A, B, C], [D, E] ]

Note that resulting arrays can have different length, and n-way splits are possible. A strided split of the same array looks thus (rightmost example of Figure 2):

**split**( [A, B, C, D, E], 2, **true** ) → [ [A, C, E], [B, D] ]

Because of the absence of aliasing, splitting maintains the abstraction of a physical split, meaning that from the programmer's perspective, splitting behaves as if the input array was divided into disjoint subarrays. This means that an array capability is always indexed from zero, and its index range is always a consecutive range.

Logical splitting allows splitting operations on large arrays to be implemented efficiently, for example by translating (statically or dynamically) indexes used on a split capability to the correct indexes on the underlying array. For example, in the case of the consecutive split above index $i$ of the second array translates to $i+3$ on the underlying array. In the case of the strided split, index $i$ on the second array translates to $2 \cdot i + 1$ on the underlying array. Figure 3 shows this pictorially.

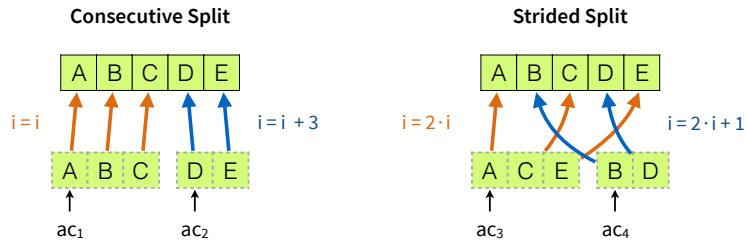

**Figure 3** Index translation between representations.

**Non-Exclusive ⇒ No Splitting** Splitting technically introduces aliases to an array, but uses partitioning to render these aliases innocuous. Array capabilities which do not give exclusive access to the underlying array can be safely (from a data-race freedom perspective) aliased without splitting as summarised by this table:

| Access Mode | Aliasing is safe because… |
| --- | --- |
| Local | All new array capabilities are local to the current thread |
| Read | No mutation is allowed |
| Locked | Concurrent access to the same data is guarded by locks |

## 3.5 Merging Array Capabilities

Merging allows the creation of new array capabilities governing access to the combined resources of its constituent parts. Merge behaves as the dual of split.

Expressed as a function, the merge operation takes the following form:

**merge**( Array[Array[t]], concat : **bool** ) : Array[t]

It takes as argument an ordered sequence of array capabilities (an array of arrays) and stitches them together, either by concatenating them or by interleaving them.

Concatenating the result of the strided split in figure 2 reorders the elements:

**merge**( [ [A, C, E], [B, D] ], **true** ) → [A, C, E, B, D]

Using interleaving, merge restores the initial state:

**merge**( [ [A, C, E], [B, D] ], **false** ) → [A, B, C, D, E]

Interleaving arrays with lengths $A$ and $B$ interleaves the $min(A,B)$ first elements of the arrays and concatenates the remaining $max(A,B) - min(A,B)$ elements from the longer array.

Just like split, merge works at the capability-level, meaning that merge does not affect the underlying array and that element indexes must be translated to the corresponding index of the underlying array. In the case of the 2nd merge example above, the logical and physical representations coincide, meaning that after the merge, no index translation is necessary. In the case of the first merge example, the index translations are still necessary (and can be derived using composition):

$$translate(i) = \begin{cases} 2 \cdot i & i < 3 \\ 2 \cdot (i-3) + 1 & i \geq 3 \end{cases}$$

Merging at the capability-level requires exclusive capabilities or immutable arrays—otherwise, changes to overlapping capabilities would be visible.

### 3.6 Aligning Logical and Physical Representation

Expressed as a function, the align operation takes the following form:

$$\textbf{align}(\ \mathsf{Array[t]}\ )\ :\ \mathsf{Array[t]}$$

From the semantics of the program, align is the identity function. Under the hood, however, it aligns the logical and physical representation by altering the latter in accordance with the former.

Thus, the following call to align on the array capability from the previous section might seem useless:

$$\textbf{align}(\ [\mathsf{A,\ C,\ E,\ B,\ D}]\ )\ \rightarrow\ [\mathsf{A,\ C,\ E,\ B,\ D}]$$

However, after this call, the index translation function of the resulting array is simply $translate(i) = i$ which clearly means faster accesses, but more importantly facilitates operating on an array in a cache-friendly access pattern.

To allow aligning in constant (already allocated) space, it needs access to all parts of an array touched by the alignment operation. Therefore align may only be called on capabilities which have no siblings (i.e., they are not split). We leave the possibility of calling align on partial capabilities as future work.

### 3.7 Using Split, Merge and Align in Concert

To show how the operations can be fruitfully combined, we show a number of short examples. We use syntax from the Encore programming language [8], which is reminiscent of Java and Scala. The following code snippet splits an array capability into two containing all odd and even elements, then uses a foreach loop to spawn two tasks that operate on each subarray.

```
1 for c <- split( array, 2, true ) { // [1,2,3,4]   => [ [1,3],  [2,4]  ]
2   async {
3     for e <- c { call_function( e ) }
4   }
5 }
```

To operate on the last 2 elements of every group of 3, we first performs a 3-way split, then discard the first array, and merge the remaining.

```
1 var s = split( array, 3, true ) // [1,2,3,4,5,6]   => [ [1,4],  [2,5],  [3,6]  ]
2 for e <- merge( [ s[1],  s[2] ],  false ) { // [ [2,5],  [3,6]  ] => [2,3,5,6]
3   call_function( e )
4 }
```

The following code snippet applies a function to all elements of an array in parallel, divide-and-conquer-style. Like the code snippets above, the function effectively destroys the array capability a during its operation. We will revisit this issue in § 3.8.

```
1  def apply( a : Array[t], fun : t -> unit ) : unit
2    if a.length() < SEQUENTIAL_CUTOFF {
3      for e <- a { fun( e ) }
4    } else {
5      val s = split( a, 2, false )
6      async { apply( s[0], fun ) }
7      apply( s[1], fun )
8    }
9  }
```

Using split and merge, rotating a col × row matrix *logically* is simple:

$$\textbf{val } \text{rotated} = \textbf{merge}( \textbf{split}( \text{matrix}, \text{cols}, \textbf{true} ), \textbf{true} )$$

To perform the same rotation *physically*, simply wrap the operation in an **align**:

$$\textbf{val } \text{rotated} = \textbf{align}( \textbf{merge}( \textbf{split}( \text{matrix}, \text{cols}, \textbf{true} ), \textbf{true} ) )$$

Let us walk through the latter two examples step by step. Let matrix be a matrix with 2 rows and 3 columns, [ABCDEF]. The operation **split**([ABCDEF], 3, **true**) returns the nested array [ [AD], [BE], [CF] ]. Note that each inner array now holds a column of the matrix. The result can be concatenated by merge in a straightforward manner: **merge**([ [AD], [BE], [CF] ], **true**). This returns [ADBECF], which matches the logical representation of the rotated matrix in the programmer's mind.

Figure 4 shows this pictorially step by step. The top left subfigure shows the matrix from the view of the programmer; the top right shows the initial mismatch between her mental model and the physical representation. Note that initially, the logical representation of the array initial capability $ac_0$ is aligned with the physical representation. Following the split, the resulting array capabilities $ac_1$–$ac_3$ each hold the individual columns in the logical representation, but the physical representation remains the same. Following the merge, there is again only a single array capability $ac_4$, but the mismatch between the logical and physical representations remain. Even though iteration over $ac_4$ would move nicely across columns, each access would be translated internally. Finally, after aligning, the physical and logical representations are again aligned, but have now shifted from the initial column major mode to row major mode.

### 3.8 Borrowing—Simplifying Life with Capabilities

Borrowing [5] is a well-known technique for programming with unique references. Similar to *Kappa* [11], *Arr-O-Matic* uses borrowing for "automating" the reassembly of deconstructed capabilities.

Borrowing a capability creates a temporary copy of the capability (not the underlying resource). The original is hidden (buried) and the copy is qualified as borrowed which bounds it to the stack (i.e., it cannot be stored on the heap or in a global variable, etc.) The lifetime of the borrowed copy is tied to a particular lexical scope. Borrowedness is tracked through a type qualifier and splitting and merging preserves borrowedness.

At the end of a borrowing scope all temporary capabilities are guaranteed to be dead which means we can reveal the buried capability and reinstate the original

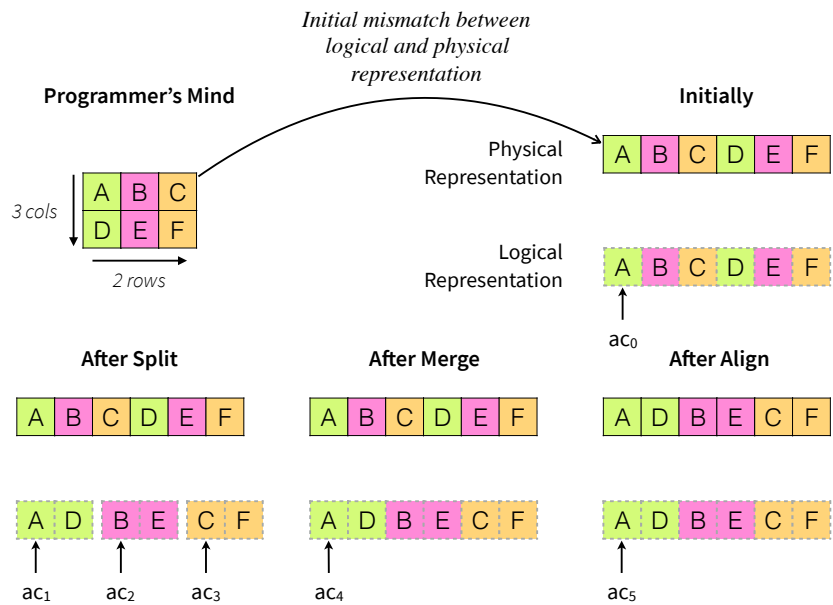

**Figure 4** Overview of the individual steps of matrix rotation from column major mode to row major mode using splitting, merging and aligning.

variable and clear the borrowed type qualifier. A borrowing block typically has a source variable, a target variable, and a scope:

**borrow** $b$ **as** $c$ **in** { ... }

Inside the ..., $c$ holds a copy of the buried original and has its type changed to reflect its borrowing status (e.g., **borrowed** T) *plus potential access mode weakenings* (as exemplified by listing 3). Once the ... exits, $c$, any aliases of $c$ (including those created by passing $c$ as arguments to function, results from splitting $c$, etc.) are unreachable to the program. This makes it safe to reinstate $b$ with the buried original value.

*Arr-O-Matic*'s array capabilities support borrowing and requires splitting and merging to preserve a capability's borrowing status. Thus, borrowing can be used to simplify many divide-and-conquer operations that rely on recursive splitting by letting them forgo merging. We exemplify this by revisiting the example in listing 2. A naive translation of that code to use reference capabilities would destroy the array because preservation of uniqueness requires that each unique capability is moved into each subtask, and no work is done to rebuild the matrix after apply() is done.

To avoid this destructive behaviour without having to add complicated logic for reassembling the matrix, we simply change the apply function's arguments to **borrowed read** and **borrowed** respectively to capture that they are borrowed. As a borrowed reference ultimately is a copy of a unique reference that can be reinstated, and because splitting preserves borrowing, we can safely destroy the to array, finally calling the same do_row() as in listing 2, because the original matrix will be reassembled after the borrowing ceases.

Lines 12–19 of listing 2 illustrate two uses of borrowing. Each iteration in the while loop corresponds to one phase in the HAParaNDA algorithm. First, from is borrowed

**Reference Capabilities for Safe Parallel Array Programming**

■ **Listing 3** Revisiting the code in listing 2 using borrowing and splitting in concert.

```
1  // Updated apply from listing 2
2  def apply( from : borrowed read Array[int], to : borrowed Array[int], rows : int,  row : int ) :
       ↪ unit
3    if rows > 1 {
4      var sub = split( to, 2, false ) // Split matrix in half
5      finish { async apply( from, sub[0], ( rows + 1 ) / 2, row )
6                async apply( from, sub[1], rows / 2, row + rows / 2 ) } }
7    } else {
8      do_row( from, to, row )
9    }
10 }
11 // Using borrowing to call apply
12 while ( some_condition ) {
13   borrow from as from_ : [val int] in { // borrow into read-only
14     borrow to as to_ in {
15       apply( from_, to_, rows, 0 )
16     }
17   }
18   val tmp = from; from = to;  to = tmp // flip to and from
19 }
```

into an aliasing from_ whose access mode is **read**. This weakening allows us to share the read-only matrix between all parallel tasks created by apply, but after the borrowing ceases, the matrix is again mutable which allows us to flip from and to. The borrowing of to into the aliasing to_ allows us to repeatedly split to_ inside the array, but as the resulting capabilities are all inaccessible when the borrowing ceases, we can safely reinstate to again, with access to the entire matrix. Similar tricks avoid destructive behaviour in the examples in section 3.7.

## 4 Formalism

We investigate our design in a core calculus that does away with most things except array operations and basic concurrency operations. We omit control structures like conditionals and loops, but notably support recursion. These simplifications are standard and without essential loss of generality in order to not detract from the main point of interest for this paper. We focus our calculus on array capabilities with the unique and read-only access modes (cf. 3.3). The locked and local modes are also useful, but less interesting to study as their safety in a concurrent setting is more obvious: local arrays will never be shared between units of concurrency, and accesses to locked arrays are always synchronised using locks. We also omit operations related to the physical representation of arrays (e.g., aligning).

We prove the calculus sound, and show that evaluation preserves *array disjointness*: any two coexisting array capabilities either refer to different arrays, refer to non-overlapping segments of the same array, or are both read-only capabilities. A corollary of this property is the absence of data-races. We model concurrency using structured threads: a thread can spawn two threads running arbitrary expressions, wait for them to finish, and then continue running. Spawned threads can pass values to the parent thread by writing to arrays. There is no way for two sibling threads to communicate values (as shared access to mutable state would be considered a data-race).

**Arrays and Array Operations**   Freshly created arrays are **unique**, meaning there is only one single reference to that array in the system. Uniqueness is preserved through destructive reads [7, 20]. For simplicity, destructive reads are implicit. Other means of preserving uniqueness are possible, such as alias burying [5], but this is orthogonal to operations on unique arrays, which is our focus here.

Uniqueness may be relaxed through borrowing, which is standard. Borrowing allows an array to be aliased under a new name while the old name is *buried*—temporarily made inaccessible. This allows relaxed treatment of the borrowed alias knowing that once the borrowing ceases, all aliases are out of scope and the uniqueness of the original reference can be restored. This allows typical patterns like divide-and-conquer style recursive splitting of an array, without having to explicitly manage the merging; one can just fall back to the original reference once the borrowing ceases.

We model the read access mode by annotating array types as mutable (**var**) or immutable (**val**); an array capability with the **val** annotation can be copied as long as it does not contain unique capabilities itself, but cannot be updated. Because we use destructive reads to preserve uniqueness of mutable arrays, we also cannot extract a **var** array element from a **val** array. Relaxations of this are possible (e.g., reverse borrowing [11]). We allow temporarily turning a **var** array into a **val** array during borrowing.

Splitting an array also introduces aliases, *but only safe ones*, since aliases give access to disjoint parts of the underlying array. The disjointness comes from explicated index translation functions, as introduced in § 3, which are denoted $\sigma$ in our calculus. Each $\sigma$ is a function **int** $\to$ **int** that maps an index in an array capability to an index in the original array. Each array alias carries its own index translation function around as a subscript.

Whenever an array capability is split, we require that the index translation functions for the introduced aliases do not remap accesses to the same index in the original array, and maintain the invariant that the introduced aliases cannot be used to access elements that were not accessible in the original capability. This way, we do not tie the calculus to a specific set of possible splits.

Synthesising index translation functions for consecutive and strided splitting (cf. § 3) that satisfy this invariant is straightforward. For a newly created array of length $2 \cdot n$, its index translation function is the identity function:

$$\sigma = \{(0,0),(1,1),\ldots,(2 \cdot n - 1, 2 \cdot n - 1)\}$$

A consecutive split of this array yields the following two translation functions:

$$\sigma_1 = \{(0,0),(1,1),\ldots,(n-1,n-1)\} \qquad \sigma_2 = \{(0,n),(1,n+1),\ldots,(n-1,2 \cdot n - 1)\}$$

**Reference Capabilities for Safe Parallel Array Programming**

If we split it in a strided fashion, we obtain the following two translation functions:

$$\sigma_1 = \{(0,0),(1,2),\ldots,(n-1,2\cdot n-2)\} \qquad \sigma_2 = \{(0,1),(1,3),\ldots,(n-1,2\cdot n-1)\}$$

We note that $[0,\mathrm{dom}(\sigma)-1]$ always denotes the range of valid indexes. We could generalise this notion and allow index translation functions which "forget" indexes accessible in the source when splitting. We also note that the size of the domains of two translation functions in a split do not need to be of the same size.

Two array capabilities with access to the same array can be merged. Since array capabilities are simply aliases of the same array with different index translation functions, a simple identity test suffices to maintain soundness. The result of a merge becomes an index translation function that is the (disjoint) union of the merged views. For simplicity, we only model concatenating merges, as interleaved merges can be recreated using additional splitting and concatenations.

**Syntax**   The syntax of *Arr-O-Matic* is unsurprising. A program consists of zero or more function declarations. For simplicity, functions take a single argument and produce a single return value. Types are either arrays or booleans. Array elements have modifiers—**var** and **val**—that capture whether they are mutable or not. We track whether an array is unique, borrowed or buried in its type.

Expressions are standard: variables, values, function calls and standard let-bindings (on the first line), array operations—reading, updating, splitting, merging and creating new arrays (on the second line), and concurrency and borrowing (on the last line).

$$
\begin{array}{rcll}
P & ::= & \overline{F} & \textit{Program} \\
F & ::= & \mathbf{fun}\ f(x:t):t'\ e & \textit{Function} \\
t & ::= & \alpha\,[mod\ t]\ |\ \mathbf{bool} & \textit{Type} \\
\alpha & ::= & \mathbf{unique}\ |\ \mathbf{borrowed}\ |\ \mathbf{buried} & \textit{Annotation} \\
mod & ::= & \mathbf{var}\ |\ \mathbf{val} & \textit{Modifier} \\
e & ::= & x\ |\ v\ |\ f(e)\ |\ \mathbf{let}\ x=e\ \mathbf{in}\ e' & \textit{Expressions} \\
& | & x[i]\ |\ x[i]=e\ |\ \mathbf{let}\ y_\sigma \uplus z_\sigma = x\ \mathbf{in}\ e\ |\ x \uplus y\ |\ \mathbf{new}\ t(i) & \textit{(arrays)} \\
& | & \mathbf{finish}\{\mathbf{async}\{e_1\}\ \mathbf{async}\{e_2\}\};e_3\ |\ \mathbf{borrow}\ x\ \mathbf{as}\ y\ \mathbf{in}\ e\ |\ \boxed{\mathbf{B}(e)} & \textit{(conc.)}
\end{array}
$$

The only value that appears in the static semantics is **null**. We introduce run-time values in the dynamic semantics later. The expression $\mathbf{B}(e)$ is used to track the scope of borrowing, and only appears during evaluation.

### 4.1 Static Semantics

Expressions are typed under an environment, $\Gamma$, that maps variables to types ($x:t$). Nested scopes introduced by borrowing use • (as opposed to a comma) and always require the presence of the borrowed variable earlier in the environment. This allows us to identify which buried variable to restore after borrowing ends. The type rules also use an additional environment $\Delta$ which maps locations of arrays to their types. It is always empty during static checking, but is used at run-time to check for well-formedness of configurations.

$$
\begin{array}{ccc}
\text{WF-ENV} & \text{WF-VAR} & \text{WF-BLOCK} \\
& \vdash \Gamma \quad \vdash t \quad x \notin \text{vars}(\Gamma) & \vdash \Gamma \quad \vdash t \quad x \in \text{vars}(\Gamma) \\
\hline
\vdash \epsilon & \vdash \Gamma, x : t & \vdash \Gamma \bullet x : t
\end{array}
$$

The only basic type is **bool**. Any well-formed type can be used as elements in an array type. A well-formed program consists of well-formed functions. We assume the presence of a function main which is used as the starting point of the program (and which ignores its parameter). We use the notation $\overline{P_i}^n$ for the sequence $P_1 \ldots P_n$.

$$
\begin{array}{cccc}
\text{WF-BOOL} & \text{WF-ARRAY} & \text{WF-PROGRAM} & \text{WF-FUNCTION} \\
& \vdash t & \overline{\vdash F_i}^n & \vdash t \quad \vdash t' \quad \Gamma, x : t \vdash e : t' \\
\hline
\vdash \textbf{bool} & \vdash \alpha \, [mod \, t] & \vdash F_1 .. F_n & \vdash \textbf{fun} \, f(x : t) : t' \, e
\end{array}
$$

Arrays are at the heart of the system. By rule E-ARRAY-NEW, any type can be used to create a unique array of $n$ elements. By rule E-ARRAY-LOOKUP, we can look up elements from array-typed variables. If the elements are not read-only values (booleans or arrays with **val** modifiers at all levels), the elements must be destructively read to preserve uniqueness, which requires that the modifier of the array is **var**. If the elements are read-only values, they can be read non-destructively. By rule E-ARRAY-ASSIGN, we may update elements of array-typed variables. This requires that the array is mutable (**var**) and that the element stored is not **borrowed**. Because we do not model a **void** or **unit** type, assignments to existing array elements (within bounds) return **true**.

$$
\begin{array}{cc}
\text{E-ARRAY-NEW} & \text{E-ARRAY-LOOKUP} \\
\vdash \Delta \quad \vdash \Gamma \quad \vdash t & \vdash \Gamma \quad \vdash \Delta \quad \Delta; \Gamma \vdash x : \alpha \, [mod \, t] \\
t = \textbf{unique} \, [mod \, t'] & \neg \, \text{readOnly}(t) \Rightarrow mod = \textbf{var} \\
\hline
\Delta; \Gamma \vdash \textbf{new} \, t(n) : t & \Delta; \Gamma \vdash x[i] : t
\end{array}
$$

$$
\begin{array}{c}
\text{E-ARRAY-ASSIGN} \\
\Delta; \Gamma \vdash e : t \quad \Delta; \Gamma \vdash x : \alpha \, [\textbf{var} \, t] \quad \neg \, \text{borrowed}(t) \\
\hline
\Delta; \Gamma \vdash x[i] = e : \textbf{bool}
\end{array}
$$

Values of array type can be split into equi-typed subarrays (E-ARRAY-SPLIT) and merged back (E-ARRAY-MERGE). The index translation functions for the array's subarrays must always map to disjoint indexes (*i.e.*, their ranges must be disjoint). Finally, by rule E-BORROW, we may borrow arrays onto the stack under a new name, while shadowing the original, making it temporarily buried (not accessible to expressions), and the borrowed alias optionally read-only (R($t$) recursively changes all modifiers of an array type to **val**).

$$
\begin{array}{cc}
\text{E-ARRAY-SPLIT} & \\
\Delta; \Gamma \vdash x : t \quad \Delta; \Gamma, y : t, z : t \vdash e : t_1 & \text{E-ARRAY-MERGE} \\
\text{array}(t) \quad \text{rng}(\sigma_1) \cap \text{rng}(\sigma_2) = \emptyset & \Delta; \Gamma \vdash x : t \quad \Delta; \Gamma \vdash y : t \quad \text{array}(t) \\
\hline
\Delta; \Gamma \vdash \textbf{let} \, y_{\sigma_1} \uplus z_{\sigma_2} = x \, \textbf{in} \, e : t_1 & \Delta; \Gamma \vdash x \uplus y : t
\end{array}
$$

$$
\begin{array}{c}
\text{E-BORROW} \\
\Delta; \Gamma \vdash x : \alpha \, [mod \, t'] \quad \Delta; \Gamma \bullet x : \textbf{buried} \, [mod \, t'], y : t'' \vdash e : t \\
t'' = \textbf{borrowed} \, [mod \, t'] \vee t'' = \textbf{borrowed} \, [\textbf{val} \, R(t')] \\
\hline
\Delta; \Gamma \vdash \textbf{borrow} \, x \, \textbf{as} \, y \, \textbf{in} \, e : t
\end{array}
$$

**Reference Capabilities for Safe Parallel Array Programming**

In addition to the array operations, we have standard **let** bindings (E-LET), lookup of non-buried variables (E-VAR) and function call (E-CALL).

$$
\begin{array}{ccc}
\text{E-LET} & \text{E-VAR} & \text{E-CALL} \\
\dfrac{\Delta;\Gamma \vdash e_1 : t_1 \quad \Delta;\Gamma, x:t_1 \vdash e_2 : t}{\Delta;\Gamma \vdash \textbf{let}\, x = e_1 \,\textbf{in}\, e_2 : t} & \dfrac{\vdash \Delta \quad \vdash \Gamma \quad \Gamma(x) = t \quad \neg\,\text{buried}(t)}{\Delta;\Gamma \vdash x : t} & \dfrac{P(f) = \textbf{fun}\, f(x:t) : t'\ e \quad \Delta;\Gamma \vdash e : t}{\Delta;\Gamma \vdash f(e) : t'}
\end{array}
$$

We use finish/async blocks to allow one thread to spawn two additional threads which run to completion before the spawning thread continues with another expression. We require that the two **async** blocks do not touch the same variables (fv($e$) extracts the free variables in $e$). This is essentially a Separation logic-style frame rule, except that we assert a relation directly between the asynchronous expressions, as opposed to partitioning the environment under which they are typed. In combination with unique and read capabilities, this guarantees data-race freedom (cf. p. 4.3)

$$
\text{E-FINISH-ASYNC} \quad \dfrac{\Delta;\Gamma \vdash e_1 : t_1 \quad \Delta;\Gamma \vdash e_2 : t_2 \quad \text{fv}(e_1) \cap \text{fv}(e_2) = \emptyset \quad \Delta;\Gamma \vdash e_3 : t}{\Delta;\Gamma \vdash \textbf{finish}\, \{\textbf{async}\, \{e_1\}\, \textbf{async}\, \{e_2\}\}; e_3 : t}
$$

Boolean values are **true** and **false**. The **null** value can have any array type.

$$
\begin{array}{cc}
\text{E-BOOL} & \text{E-NULL} \\
\dfrac{\vdash \Delta \quad \vdash \Gamma \quad v \in \{\textbf{true}, \textbf{false}\}}{\Delta;\Gamma \vdash v : \textbf{bool}} & \dfrac{\vdash \Delta \quad \vdash \Gamma \quad \vdash t \quad \text{array}(t)}{\Delta;\Gamma \vdash \textbf{null} : t}
\end{array}
$$

### 4.2 Dynamic Semantics

The dynamic semantics are mostly straightforward. For brevity, we highlight two kinds of rules: operations on arrays and operations related to concurrency. The remaining rules can be found in appendix A. Run-time configurations take the form $\langle H, A \rangle$ where $H$ is the global heap storing arrays, and $A$ is a tree of activities (threads) currently running in the system.

$$
\begin{array}{rcll}
H & ::= & \epsilon \mid H, \iota \mapsto [v_1..v_n] & \textit{Heap} \\
A & ::= & (S, e) \mid A_1 \parallel A_2 \triangleright (S, e) \mid \textbf{Error} & \textit{Activities} \\
S & ::= & \epsilon \mid S, x \mapsto v \mid S \bullet x \mapsto v & \textit{Stack} \\
v & ::= & \iota_\sigma \mid \textbf{null} \mid \textbf{true} \mid \textbf{false} & \textit{Values} \\
E[\bullet] & ::= & \textbf{let}\, x = \bullet \,\textbf{in}\, e \mid x[i] = \bullet \mid f(\bullet) \mid \textbf{B}(\bullet) & \textit{Evaluation Context}
\end{array}
$$

The heap $H$ maps identities ($\iota$) to arrays. An activity $A$ is either a single thread with a stack of its own and an expression being executed, two activities being run in parallel while a single thread waits for them to finish, or an error state. The stack $S$ maps variable names to values, and may include • symbols to identify the start of a borrowing block. The value $\iota_\sigma$ is a reference to an array with identity $\iota$ and translation function $\sigma$. If $x$ maps to $\iota_\sigma$, $x[i]$ will evaluate to the value at index $\sigma(i)$ in the array identified by $\iota$. For a freshly created array, $\sigma$ is the identity function. The evaluation context $E$ is used to abstract over evaluation of subexpressions.

DYN-CONTEXT
$$\frac{\langle H,(S,e)\rangle \to \langle H',(S',e')\rangle}{\langle H,(S,E[e])\rangle \to \langle H',(S',E[e'])\rangle}$$

Splitting an array capability (DYN-ARRAY-SPLIT) consumes the original capability and introduces two aliases with access to disjoint parts of the array (captured by $\text{rng}(\sigma_1) \cap \text{rng}(\sigma_2) = \emptyset$ in the type rule E-ARRAY-SPLIT). We make sure that the two new capabilities can not access elements that were not accessible via the original capability by composing the translation functions $\sigma_1$ and $\sigma_2$ with the original translation function $\sigma$. For simplicity, we assume that the $\sigma$'s are the smallest possible translation functions, which allows their use for bounds checking at run-time.

DYN-ARRAY-NEW
$$\frac{(t = \textbf{bool} \Rightarrow v = \textbf{false}) \quad (\text{array}(t) \Rightarrow v = \textbf{null}) \quad \text{fresh}(\iota) \quad \sigma = id}{\langle H,(S,\textbf{new unique }[mod\ t](n))\rangle \to \langle H, \iota \mapsto [\overline{v_i}^n], (S, \iota_\sigma)\rangle}$$

DYN-ARRAY-SPLIT
$$\frac{S(x) = \iota_\sigma \quad \sigma'_1 = \sigma \circ \sigma_1 \quad \sigma'_2 = \sigma \circ \sigma_2 \quad S' = S[x \mapsto \textbf{null}], y \mapsto \iota_{\sigma'_1}, z \mapsto \iota_{\sigma'_2}}{\langle H,(S,\textbf{let } y_{\sigma_1} \uplus z_{\sigma_2} = x \textbf{ in } e)\rangle \to \langle H,(S',e)\rangle}$$

Merging two array capabilities consumes them and introduces a capability with the union of the original translation functions.

DYN-ARRAY-MERGE
$$\frac{S(x) = \iota_{\sigma_1} \quad S(y) = \iota_{\sigma_2} \quad S' = S[x \mapsto \textbf{null}][y \mapsto \textbf{null}]}{\langle H,(S, x \uplus y)\rangle \to \langle H,(S', \iota_{\sigma_1 \uplus \sigma_2})\rangle}$$

Accesses and updates to an array require that the index is in the domain of the translation function $\sigma$. Accessing elements whose type is unique requires that the element is nullified as a side-effect to preserve uniqueness (note that the check readOnlyElems could be done statically).

DYN-ARRAY-ASSIGN
$$\frac{S(x) = \iota_\sigma \quad i \in \text{dom}(\sigma) \quad H' = H[\iota \mapsto (H(\iota)[\sigma(i) \mapsto v])]}{\langle H,(S, x[i] = v)\rangle \to \langle H',(S, \textbf{true})\rangle}$$

DYN-ARRAY-LOOKUP
$$\frac{\text{readOnlyElems}(x) \quad S(x) = \iota_\sigma \quad i \in \text{dom}(\sigma) \quad v = H(\iota)[\sigma(i)]}{\langle H,(S, x[i])\rangle \to \langle H,(S, v)\rangle}$$

DYN-ARRAY-LOOKUP-UNIQUE
$$\frac{\neg\,\text{readOnlyElems}(x) \quad S(x) = \iota_\sigma \quad i \in \text{dom}(\sigma) \quad v = H(\iota)[\sigma(i)]}{\langle H,(S, x[i])\rangle \to \langle H[\iota \mapsto (H(\iota)[\sigma(i) \mapsto \textbf{null}])], (S, v)\rangle}$$

The system reaches an error state if an array is accessed out-of-bounds or through a **null**-reference, or if two merged array capabilities refer to different arrays.

DYN-ARRAY-ASSIGN-FAIL
$$\frac{S(x) = \iota_\sigma \quad i \notin \text{dom}(\sigma)}{\langle H,(S, x[i] = v)\rangle \to \langle H, \textbf{Error}\rangle}$$

DYN-ARRAY-LOOKUP-FAIL
$$\frac{S(x) = \iota_\sigma \quad i \notin \text{dom}(\sigma)}{\langle H,(S, x[i])\rangle \to \langle H, \textbf{Error}\rangle}$$

DYN-ARRAY-MERGE-FAIL
$$\frac{S(x) = \iota_{\sigma_1} \quad S(y) = \iota'_{\sigma_2} \quad \iota \neq \iota'}{\langle H,(S, x \uplus y)\rangle \to \langle H, \textbf{Error}\rangle}$$

DYN-ARRAY-ASSIGN-NULL
$$\frac{S(x) = \textbf{null}}{\langle H,(S, x[i] = v)\rangle \to \langle H, \textbf{Error}\rangle}$$

**Reference Capabilities for Safe Parallel Array Programming**

$$\frac{\text{DYN-ARRAY-LOOKUP-NULL}}{S(x) = \textbf{null}}$$
$$\overline{\langle H, (S, x[i]) \rangle \rightarrow \langle H, \textbf{Error} \rangle}$$

$$\frac{\text{DYN-ARRAY-MERGE-NULL}}{S(x) = \textbf{null} \vee S(y) = \textbf{null}}$$
$$\overline{\langle H, (S, x \uplus y) \rangle \rightarrow \langle H, \textbf{Error} \rangle}$$

$$\frac{\text{DYN-ARRAY-SPLIT-NULL}}{S(x) = \textbf{null}}$$
$$\overline{\langle H, (S, \textbf{let } y_{\sigma_1} \uplus z_{\sigma_2} = x \textbf{ in } e) \rangle \rightarrow \langle H, \textbf{Error} \rangle}$$

A borrowing block evaluates to a special dynamic form $\textbf{B}(e)$ which denotes that $e$ is executing inside a borrowing block. We mark the starting position of the borrowing in $S$ using the $\bullet$ marker. When the block finishes, we simply remove all the entries to the right of the last $\bullet$ in $S$—these variables will never be touched again since they were introduced in a scope that has now been reduced to a value. Because we are reinstating the borrowed value, we also discard the result of the borrowing block (in case it is the borrowed value) by replacing it with $\textbf{true}$.

$$\frac{\text{DYN-BORROW}}{S(x) = v \quad S' = S \bullet x \mapsto \textbf{null}, y \mapsto v}$$
$$\overline{\langle H, (S, \textbf{borrow } x \textbf{ as } y \textbf{ in } e) \rangle \rightarrow \langle H, (S', \textbf{B}(e)) \rangle}$$

$$\frac{\text{DYN-BORROW-DONE}}{S = S' \bullet S'' \quad \bullet \notin S''}$$
$$\overline{\langle H, (S, \textbf{B}(v)) \rangle \rightarrow \langle H, (S', \textbf{true}) \rangle}$$

Evaluating a finish/async-block results in two new threads being spawned. They are initialised with stacks containing only the variables accessed in each expression (note in a well-formed program, these variables are known to be disjoint).

$$\frac{\text{DYN-SPAWN}}{S_1 = [x \mapsto v \mid x \in \text{fv}(e_1) \wedge S(x) = v] \quad S_2 = [x \mapsto v \mid x \in \text{fv}(e_2) \wedge S(x) = v]}$$
$$\overline{\langle H, (S, \textbf{finish } \{ \textbf{async } \{e_1\} \textbf{ async } \{e_2\} \}; e_3) \rangle \rightarrow \langle H, (S_1, e_1) \parallel (S_2, e_2) \triangleright (S, e_3) \rangle}$$

$$\frac{\text{DYN-SPAWN-CTX}}{\langle H, (S, e) \rangle \rightarrow \langle H', A_1 \parallel A_2 \triangleright (S, e') \rangle}$$
$$\overline{\langle H, (S, E[e]) \rangle \rightarrow \langle H', A_1 \parallel A_2 \triangleright (S, E[e']) \rangle}$$

We model concurrency through non-deterministic choice when evaluating parallel activities. When both parallel activities have finished, the blocking thread continues.

$$\frac{\text{DYN-SCHED-L}}{\langle H, A_1 \rangle \rightarrow \langle H', A_1' \rangle}$$
$$\overline{\langle H, A_1 \parallel A_2 \triangleright (S, e_3) \rangle \rightarrow \langle H', A_1' \parallel A_2 \triangleright (S, e_3) \rangle}$$

$$\frac{\text{DYN-SCHED-R}}{\langle H, A_2 \rangle \rightarrow \langle H', A_2' \rangle}$$
$$\overline{\langle H, A_1 \parallel A_2 \triangleright (S, e_3) \rangle \rightarrow \langle H', A_1 \parallel A_2' \triangleright (S, e_3) \rangle}$$

$$\frac{\text{DYN-FINISH}}{}$$
$$\overline{\langle H, (S_1, v_1) \parallel (S_2, v_2) \triangleright (S, e_3) \rangle \rightarrow \langle H, (S, e_3) \rangle}$$

**Well-Formedness** A well-formed configuration consists of a well-formed heap and a well-formed tree of activities. A well-formed heap can be extended by a well-formed array, i.e., whose elements correspond to its type in the run-time environment $\Delta$. We also require that the domain of $\Delta$ is the same as the domain of the heap.

$$\frac{\text{WF-CFG}}{\Delta \vdash H \quad \Delta; \Gamma \vdash A : t \quad \text{dom}(\Delta) = \text{dom}(H)}$$
$$\overline{\Delta; \Gamma \vdash \langle H, A \rangle : t}$$

$$\frac{\text{WF-H-EMPTY}}{\vdash \Delta}$$
$$\overline{\Delta \vdash \epsilon}$$

$$\frac{\text{WF-H-ADD}}{\Delta \vdash H \quad \Delta(\iota) = \alpha \ [mod \ t] \quad \overline{\Delta \vdash v_i : t}^n}$$
$$\overline{\Delta \vdash H, \iota \mapsto [v_1 .. v_n]}$$

A well-formed thread has a well-formed stack and a well-formed expression. In the case of parallel activities, we require the existence of new type environments to match the stacks of the underlying threads.

$$\frac{\text{WF-THREAD}}{\Delta; \Gamma \vdash S \quad \Delta; \Gamma \vdash e : t}{\Delta; \Gamma \vdash (S, e) : t} \qquad \frac{\text{WF-ASYNC}}{\exists \Gamma_1 \,.\, \Delta; \Gamma_1 \vdash A_1 : t_1 \quad \exists \Gamma_2 \,.\, \Delta; \Gamma_2 \vdash A_2 : t_2 \quad \Delta; \Gamma \vdash S \quad \Delta; \Gamma \vdash e : t}{\Delta; \Gamma \vdash A_1 \parallel A_2 \triangleright (S, e) : t}$$

A well-formed stack may be extended, either by a new variable whose type is consistent with that in the type environment, or by a nullified variable of buried type. This allows us to identify the "scope" of a borrowing block.

$$\frac{\text{WF-S-VAR}}{\Delta \vdash v : t \quad \Delta; \Gamma \vdash S}{\Delta; \Gamma, x : t \vdash S, x \mapsto v} \qquad \frac{\text{WF-S-BORROW}}{\Delta; \Gamma \vdash S \quad \Delta; \Gamma \vdash x : \alpha \,[mod\, t]}{\Delta; \Gamma \bullet x : \text{buried}\,[mod\, t] \vdash S \bullet x \mapsto \textbf{null}}$$

### 4.3 Meta-Theoretic Properties

We prove type soundness through the usual progress and preservation scheme. Full proofs can be found in appendix B.

**Theorem 1** (Progress).
*If $\Delta; \Gamma \vdash cfg : t$ then either the program is done ($cfg = \langle H, (S, v) \rangle$), is an error state ($cfg = \langle H, \textbf{Error} \rangle$) or there exists some $cfg'$ such that $cfg \rightarrow cfg'$.*

*Proof sketch.* Proven by induction over the shape of *A* (and any *e* therein). □

**Theorem 2** (Preservation).
*If $\Delta; \Gamma \vdash cfg$ and $cfg \rightarrow cfg'$, then $\exists \Delta', \Gamma'$ such that $\Delta'; \Gamma' \vdash cfg'$ and $\Delta \subseteq \Delta'$*

*Proof sketch.* Proven by induction over the shape of *A* (and any *e* therein). The most interesting case is that of the borrowing block, where the • marker is used to discard parts of the stack, making $\Gamma'$ smaller than $\Gamma$. Similarly, the inductive cases require reasoning about how the evaluation of a borrowing block in a subexpression cannot affect the well-formedness of the full expression, as only variables that are out of scope are discarded. □

The key property of *Arr-O-Matic* however is that evaluation preserves *array disjointness*, namely that any two non-buried array capabilities in the system either refer to different arrays, refer to disjoint parts of the same array, or can only be used for reading. The property arrayDisjointness(*cfg*) is defined by gathering all array capabilities in the system (non-buried variables on the stacks of the currently executing threads, array elements on the heap, and values $\iota_\sigma$ in expressions) together with their types, and requiring that any two capabilities fulfil one of the three properties above.

Assume caps(*cfg*) extracts a set of all accessible array capabilities *c* in a configuration *cfg*. We use $c.\iota$, $c.\sigma$, and $c.t$ respectively to extract the array identifier, the index translation function, and the type of an array capability.

$$\begin{aligned}&\mathsf{arrayDisjointness}(\mathit{cfg}) \equiv \\ &\quad \forall c_1, c_2 \in \mathsf{caps}(\mathit{cfg})\ . \\ &\quad\quad c_1.\iota \neq c_2.\iota \ \lor\ \mathsf{rng}(c_1.\sigma) \cap \mathsf{rng}(c_2.\sigma) = \emptyset \ \lor\ \mathsf{read}(c_1.t) \land \mathsf{read}(c_2.t)\end{aligned}$$

**Theorem 3** (Preservation of Array Disjointness). *If $\Delta; \Gamma \vdash \mathit{cfg}$ and $\mathsf{arrayDisjointness}(\mathit{cfg})$ $\mathit{cfg} \to \mathit{cfg}'$, then $\mathsf{arrayDisjointness}(\mathit{cfg}')$*

*Proof sketch.* Proven by induction over the shape of *A* (and any *e* therein). For each case, we show that any new capabilities introduced do not overlap with an existing, mutable capability, either due to destructive reads, borrowing, or splitting. For example, in the case of a borrowing block, the buried variable is reinstated at the same time as any borrowed aliases are discarded from the stack. □

**Corollary: Data-Race Freedom** For two threads $T_1$ and $T_2$ to be able to race on the same data, they each need a capability, $c_1$ and $c_2$ respectively, such that at least one of $c_1$ and $c_2$ allows mutation, and $\mathsf{rng}(c_1.\sigma) \cap \mathsf{rng}(c_2.\sigma)$ is not empty. *Thus, preservation of array disjointness implies that parallel operations on arrays in Arr-O-Matic are guaranteed to be data-race free.*

## 5 Preliminary Evaluation

To evaluate the design of our array capabilities, we have encoded several examples and type checked them manually. We are currently exploring API design in a prototype implementation in Python, which we will report on in the future.

In brief, and pending a more thorough evaluation, our preliminary findings is that typical array algorithms that either operate on an entire array from start to finish, or break it into smaller subarrays are straightforward to implement.

**Quicksort and Merge Sort** We have encoded the typical recursive divide-and-conquer implementations of these well-known sorting algorithms. This is straightforward—the split operation works well for these kinds of algorithms, where each step further subdivides an array. This requires (and is supported "natively" by our Python prototype) the ability to split an array at a given index *i*. This can be easily achieved by the split operation in our formalism by splitting an array of length *L* into *L* pieces, then concatenating the first *i* pieces into one array and the remaining pieces into another array (but a convenience function is obviously simpler and more efficient).

Both implementations are similar to apply() in listing 3. Just like the code in that figure, both are greatly simplified by borrowing—instead of manually stitching the original array back together, we bury the original array capability before the beginning of the recursion, and simply reinstate it after the merge.

■ **Listing 4** Parallel reduction of arrays. If strided is false, adjacent elements are reduced.

```
1  // S = true => strided split with length N, else N-way adjacent split
2  def p_reduce( aa : borrowed Array[Array[int]], N : int, S : bool ) : unit {
3    finish { for a <- aa { async { reduce( a, N, S ) } } }
4  }
5
6  def reduce( a : borrowed Array[int], N : int, S : bool ) : unit {
7    focus = split( a, N, S )[0]  // Note: split(a, 1, _) = [ [a] ]
8    var sum = 0
9    for e <- focus {
10     sum += e
11   }
12   focus[0] = sum
13 }
14
15 borrow a as b in p_reduce( split( b, 8, strided ), 1, !strided )
16 borrow a as b in p_reduce( split( b, 4, strided ), 2, !strided )
17 borrow a as b in p_reduce( split( b, 2, strided ), 4, !strided )
18 borrow a as b in p_reduce( split( b, 1, strided ), 8, !strided )
```

**Stencil Application in HAParaNDA** We have encoded the stencil application of HAParaNDA, as partially shown in listing 3. In this case, borrowing is useful for matrices mediating between being read-only and shared between all threads, or mutable and private to one thread. In the latter case, splitting unique array capabilities give rise to unique subarrays.

**Parallel Reduction** As another real-world example, we looked at Harris' algorithms for parallel reduction in CUDA [18]. Harris explores several strategies for reducing—in parallel—arrays of numbers, making extensive use of unique thread IDs to ensure that no two threads access the same array elements. For example, first using $N$ threads to reduce $2 \cdot N$ elements, either adjacent or with a stride of $N$ and storing the results in the first element's place, then $N/2$ threads for summarising the result, etc. Listing 4 shows how both strategies can be encoded in *Arr-O-Matic* as one function parameterised over whether the reduced elements are adjacent or not. For clarity, we show the 4-phases of reducing a 16 element array using 8, 4, 2 and 1 threads (**async** tasks in our case) written out explicitly.

**Where *Arr-O-Matic* Falls Short** A common trick to ensure that multiple threads do not access the same elements is to use the ID of a thread in the index calculation. Subarrays through splitting solves this naturally in *Arr-O-Matic* without the need to use indices, but for this to work, splitting must happen *before* each thread is given access to its part. This means that it is not possible to delay the decision of how many parts an array should be split into until after the threads have been created. In listing 4, this is avoided because each phase creates its own set of tasks, which may be mapped to longer-running threads. A system without lightweight tasks would likely suffer starting and stopping threads.

Furthermore, *Arr-O-Matic* works best when algorithms use constant strides or splits. Iterating *diagonally* over a matrix, for example, works best in a single thread using typical index-based accesses. With multiple threads, the matrix must be aggressively split into single-element subarrays which must subsequently be distributed among the participating threads.

## 6  Related Work

*Arr-O-Matic* is an extension of Arrgh [1] which captured the initial ideas. The key property of Arrgh is similar to *Arr-O-Matic*—array disjointness—but was not stated and proven in the context of a parallel calculus. The split and merge operations in Arrgh are not fully worked out, and support only a limited form of splitting and merging. There is also no concept of **align**.

Deterministic Parallel Java (DPJ) [3] approaches similar problems, that is to allow reasoning about determinism in presence of concurrency and imperative object-orientation, by extending Java with a type and effects system. DPJ introduces index-parameterised arrays and subarrays to specifically support common parallel array algorithms. Subarrays in DPJ are implemented by `DPJArray` by wrapping a Java array and providing a view into it, with a certain start position and length. Similar to the arrays in *Arr-O-Matic*, copying is not required and an index from a subarray is translated to the index in the original array. The actual splitting of arrays is managed by the `DPJPartition`, which given pivot values where the split will be made results in a collection of `DPJArray` that are mutually disjoint subarrays of the original array. Index-parameterised arrays allow parallel operations on elements stored at different indexes of an array. By including the actual index $i$ in the type of array element at index $i$, all array elements can be distinguished from each other. One complicating factor, however, is that the compiler needs to calculate the values of any expressions used to calculate indexes, e.g., `e1` and `e2`, at compile time to be able to statically prove that `arr[e1]` and `arr[e2]` are disjoint. This is not generally possible and thus delays detection of race conditions to run-time. This problem will never occur in *Arr-O-Matic* due to the reference capabilities of *Kappa*. Two accesses from different parts of the program will never touch the same array element.

Pony [13] is an actor-based object-oriented, high-performance programming language using reference capabilities to make sure that programs are free from data-races and deadlocks. Pony arrays are treated in the same way as other objects and the declared capability will be used to control sharing between different threads of execution and mutation is only allowed if the array is not shared. There is however no support for splitting and merging like the support built into *Arr-O-Matic*.

Rust [27] avoids data-races by only allowing mutation of unaliased data. Creating subarrays can be done in Rust by slicing, and the resulting slice will be a borrowed section of the array, but there can only be one slice per array at a time without resorting to "unsafe"-blocks. Libraries using unsafe allow reusing Rust's borrowing mechanism to prevent data races, even when arrays are used in parallel code. This means that although Rust does not use array capabilities explicitly, similar properties can be emulated within the language (albeit using unsafe code).

Futhark [19] is a functional array language intended for data-parallel programs with high demands on efficiency. Its uses uniqueness types to support in-place updates in arrays to avoid unnecessary expensive copying. In *Arr-O-Matic*, given its mutable object-oriented context, uniqueness and disjointness of subarrays is used to guarantee that parallel in-place updates will never lead to data-races. Futhark supports logical splitting of arrays through operators rather than manual index tracking.

Ypnos [24] is a domain-specific language embedded in Haskell for expressing parallel stencil computations over multidimensional, array-like structures. It relies on the property of "single, independent writes", a less general version of the array disjointness of *Arr-O-Matic*, to ensure that two writes which are part of a parallel stencil computation will never overlap. While Ypnos is a less general language, it arguably handles stencil computations more elegantly than *Arr-O-Matic*. Exploring a cross-fertilisation of the two languages is an interesting direction for future work.

## 7 Conclusion

In prior work reference capabilities have shown useful to guaranteeing absence of data races in object-oriented parallel programs. Reference capabilities, however, have been applied at the object level in a way that disables the use of e.g., recursive divide-and-conquer algorithms on arrays.

We have presented *Arr-O-Matic*, an extension of Arrgh and *Kappa* with a new type of reference capabilities, array capabilities, that support e.g., logical splitting and merging of arrays. We have expressed the array capabilities in a parallel calculus and also proved that they preserve *Kappa*'s guarantees of freedom from data races.

As a next step, we intend to implement array capabilities in the language Encore to allow further exploration of the array splitting and merging design space.

**Acknowledgements**   The authors wish to thank the reviewers for their helpful suggestions, including the suggestion to look at Harris' parallel reduction.

## A  Rules Missing from Main Article

This section contains the static and dynamic rules that we omitted from the main article.

### A.1  Dynamically Tracking Variable Types

In our dynamic semantics, we need to use destructive reads in order to preserve uniqueness of unique references. Instead of bloating the semantics with duplicated rules for explicit destructive reads (e.g., $x$ vs. **consume** $x$, and $x[i]$ vs. **consume** $x[i]$), we assume a simple post-processing of well-typed programs which annotates each variable with its type. This allows dynamic rules to query the type of variables and insert destructive reads when needed. In a real implementation, this could be done fully statically, as we always know the static type of a variable. This is just a way to keep the complexity of the formal system down.

The queries used in the dynamic rules are readOnly($x$) and readOnlyElems($x$). The former checks if a variable contains a read-only capability, meaning no destructive reads are needed. A capability is read-only if its type is an array-type that is **val** on all levels (note that the innermost type is always **bool** in our calculus). readOnlyElems performs the same check on the contents of an array (but allows the outermost array to be **var**).

## A.2 Omitted Static Rules

The only omitted static rules are the typing rule for the borrowing wrapper **B** and the well-formedness rules for the runtime environment $\Delta$. All of them are straightforward.

$$\frac{\Delta;\Gamma \vdash e : t}{\Delta;\Gamma \vdash \mathbf{B}(e) : t} \text{ E-BORROWED} \qquad \frac{}{\vdash \epsilon} \text{ WF-D-EMPTY} \qquad \frac{\vdash \Delta \quad \vdash t \quad \iota \notin \mathsf{dom}(\Delta)}{\vdash \Delta, \iota \mapsto \alpha[\mathit{mod}\, t]} \text{ WF-D-ADD}$$

## A.3 Omitted Dynamic Rules

We omitted several dynamic rules. Variables are read from the stack, possibly destructively (if the type of the variable requires it to preserve uniqueness).

$$\frac{\mathsf{readOnly}(x) \quad S(x) = v}{\langle H, (S, x) \rangle \to \langle H, (S, v) \rangle} \text{ DYN-VAR-LOOKUP} \qquad \frac{\neg\,\mathsf{readOnly}(x) \quad S(x) = v}{\langle H, (S, x) \rangle \to \langle H, (S[x \mapsto \mathbf{null}], v) \rangle} \text{ DYN-VAR-LOOKUP-DEST}$$

The rules for **let**, function calls and array creation are straightforward.

$$\frac{S' = S, x \mapsto v}{\langle H, (S, \mathbf{let}\, x = v\,\mathbf{in}\, e) \rangle \to \langle H, (S', e) \rangle} \text{ DYN-LET} \qquad \frac{P(f) = \mathbf{fun}\, f(x:t):t'\, e \quad S' = S, x \mapsto v}{\langle H, (S, f(v)) \rangle \to \langle H, (S', e) \rangle} \text{ DYN-CALL}$$

$$\frac{(t = \mathbf{bool} \Rightarrow v = \mathbf{false}) \quad (\mathsf{array}(t) \Rightarrow v = \mathbf{null}) \quad \mathsf{fresh}(\iota) \quad \sigma = id}{\langle H, (S, \mathbf{new\ unique}\,[\mathit{mod}\, t](n)) \rangle \to \langle H, \iota \mapsto [\,\overline{v_i}^{\,n}\,], (S, \iota_\sigma) \rangle} \text{ DYN-ARRAY-NEW}$$

Errors are propagated by the evaluation context $E$ and across threads. Finally error states are always well-formed.

$$\frac{\langle H, (S, e) \rangle \to \langle H, \mathbf{Error} \rangle}{\langle H, (S, E[e]) \rangle \to \langle H, \mathbf{Error} \rangle} \text{ DYN-CONTEXT-ERROR} \qquad \frac{}{\langle H, \mathbf{Error} \parallel A_2 \triangleright (S, e_3) \rangle \to \langle H, \mathbf{Error} \rangle} \text{ DYN-SCHED-L-FAIL}$$

$$\frac{}{\langle H, A_1 \parallel \mathbf{Error} \triangleright (S, e_3) \rangle \to \langle H, \mathbf{Error} \rangle} \text{ DYN-SCHED-R-FAIL} \qquad \frac{}{\Delta;\Gamma \vdash \mathbf{Error} : t} \text{ WF-ERROR}$$

## B Full Soundness Proofs

This section contains the full proofs of soundness, including proofs of lemmas. We prove type soundness through the usual progress and preservation scheme. Additionally, we prove the preservation of array disjointness, which as a corollary implies data-race freedom.

**Reference Capabilities for Safe Parallel Array Programming**

### B.1 On the Scope of Borrowed Variables

In order to track that borrowing is taking place, the dynamic semantics uses a special wrapper **B**($e$) and inserts a special marker • into the stack to mark that all variables to the right of this marker on the stack were introduced during the scope of the borrowing. When looking at an expression being evaluated, we can calculate the *borrowing depth* of a subexpression by counting the number of **B**s encountered above it. For example, in the expression **let** $x$ = **B**(**B**($e_1$)) **in** $e_2$, the subexpression $e_1$ is at borrowing depth 2, while the subexpression $e_2$ is at depth 0.

Since we only introduce • markers on the stack when introducing **B** wrappers, and only remove them when exiting **B** wrappers, it is easy to see that we will always have the same number of **B** wrappers in an expression as there are • markers on the stack. An important property that follows from this relation—and from the nested nature of our expressions—is that a free variable in any expression $e$ must appear at a borrowing depth that is greater or equal to the number of • markers to the left of that variable on the stack. This also means that a variable on the stack that occurs after $d$ • markers must exist at a borrowing depth of (at least) $d$. For example, if the expression **let** $x$ = **B**($e_1$) **in** $e_2$ is at depth $d$, and the variable $x$ appears on the stack after $d + 1$ bullets, $x$ can be a free variable in $e_1$, but not in $e_2$ (assuming that $e_2$ does not contain additional **B** wrappers).

Another important property of borrowing is that while a unique capability is borrowed, all aliases of this capability, except the original which is buried, are of **borrowed** type, and exist on the stack to the right of the • marker introduced at the start of the borrowing. To see this, consider a situation where we evaluate **borrow** $x$ **as** $y$ **in** $e$. In the resulting stack, by rule DYN-BORROW and rule E-BORROW, we introduce a • marker, a nullification of $x$ and an alias of $x$ of **borrowed** type. We have no assignment of existing variables, so any new aliases on the stack will new variables to the right of the original alias (introduced via **let** or similar constructs). By rule E-ARRAY-ASSIGN, we cannot assign a **borrowed** value to an array, so we cannot introduce aliases on the heap. Thus, when the borrowing finishes, all variables to the right of the • marker are dropped (DYN-BORROW-DONE), meaning that *all* aliases of the borrowed capability are dropped, and the uniqueness of the capability being unburied is restored.

### B.2 Progress

The progress formulation and proof are standard.

**Theorem 1** (Progress).
*If $\Delta; \Gamma \vdash cfg : t$ then either the program is done (cfg = $\langle H, (S, v) \rangle$), is in an error state (cfg = $\langle H, \textbf{Error} \rangle$) or there exists some cfg$'$ such that cfg $\rightarrow$ cfg$'$.*

Assumptions:
**wf-cfg** $\Delta; \Gamma \vdash cfg : t$
We start by induction over the shape of $A$.
**Case 1.** $cfg = \langle H, \textbf{Error} \rangle$
Progress holds trivially.

**Case 2.** $cfg = \langle H, A_1 \parallel A_2 \triangleright (S, e) \rangle$
Induction hypotheses (note that $A_1$ and $A_2$ are well-formed by *wf-cfg*):
**IH1:** $A_1 = \mathbf{Error} \lor \exists H', A_1'. \langle H, A_1 \rangle \to \langle H', A_1' \rangle \lor A_1 = (S_1, v)$
**IH2:** $A_2 = \mathbf{Error} \lor \exists H', A_2'. \langle H, A_2 \rangle \to \langle H', A_2' \rangle \lor A_2 = (S_2, v)$
We continue by case analysis over **IH1**.

**Case 2.1.** $cfg = \langle H, \mathbf{Error} \parallel A_2 \triangleright (S, e) \rangle$
The configuration steps to $\langle H, \mathbf{Error} \rangle$ by rule DYN-SCHED-L-FAIL.

**Case 2.2.** $cfg = \langle H, A_1 \parallel A_2 \triangleright (S, e) \rangle$, and $\langle H, A_1 \rangle \to \langle H', A_1' \rangle$.
The configuration steps to $\langle H', A_1' \parallel A_2 \triangleright (S, e) \rangle$ by rule DYN-SCHED-L.

**Case 2.3.** $cfg = \langle H, (S_1, v_1) \parallel A_2 \triangleright (S, e) \rangle$
We continue by case analysis over **IH2**.

**Case 2.3.1.** $cfg = \langle H, (S_1, v_1) \parallel \mathbf{Error} \triangleright (S, e) \rangle$
The configuration steps to $\langle H, \mathbf{Error} \rangle$ by rule DYN-SCHED-R-FAIL.

**Case 2.3.2.** $cfg = \langle H, (S_1, v_1) \parallel A_2 \triangleright (S, e) \rangle$, and $\langle H, A_2 \rangle \to \langle H', A_2' \rangle$.
The configuration steps to $\langle H', (S_1, v_1) \parallel A_2' \triangleright (S, e) \rangle$ by rule DYN-SCHED-R.

**Case 2.3.3.** $cfg = \langle H, (S_1, v_1) \parallel (S_2, v_2) \triangleright (S, e) \rangle$
The configuration steps to $\langle H, (S, e) \rangle$ by rule DYN-FINISH

**Case 3.** $cfg = \langle H, (S, e) \rangle$
New assumption:
*wf-e*: $\Gamma \vdash e : t$, by *wf-cfg*
We proceed by induction over the shape of $e$. To avoid cluttering the presentation, we prove all of the inductive cases (which are analogous) together in a single case. This means that in the rest of the cases, we assume that all subexpressions that can be selected using the evaluation context $E$ are values.

**Case 3.1.** $e = v$
Configuration is done. Progress holds trivially.

**Case 3.2.** $e = x$
We do a case analysis on whether the type of $x$ is read-only or not (and thus if we need to do a destructive read or not):

**Case 3.2.1** readOnly($x$)
1. $S(x) = v$, by Lemma 1 (The stack mirrors $\Gamma$) with *wf-e*.
2. By rule DYN-VAR-LOOKUP with (1), the configuration steps to $\langle H, (S, v) \rangle$.

**Case 3.2.2** $\neg$readOnly($x$)
1. $S(x) = v$, by Lemma 1 (The stack mirrors $\Gamma$) with *wf-e*.
2. By rule DYN-VAR-LOOKUP-DEST with (1), the configuration steps to $\langle H, (S[x \mapsto \mathbf{null}], v) \rangle$.

**Case 3.3.** $e = f(v)$
1. $P(f) = \mathbf{fun}\ f(x : t) : t'\ e'$ by *wf-e* (E-CALL).
2. By rule DYN-CALL with (1), the configuration steps to $\langle H, (S[x \mapsto v], e') \rangle$.

**Case 3.4.** $e = \mathbf{let}\ x = v\ \mathbf{in}\ e'$
By rule DYN-LET, the configuration steps to $\langle H, ((S, x \mapsto v), e') \rangle$.

**Case 3.5.** $e = x[i]$

1. $\Delta; \Gamma \vdash x : \alpha \ [mod \ t]$ by **wf-e** (E-ARRAY-LOOKUP).
2. $S(x) = v$, and $\Delta; \Gamma \vdash v : \alpha \ [mod \ t]$ by Lemma 1 (The stack mirrors $\Gamma$) with (1).
3. We do a case analysis on whether $v$ is **null** or a reference.
    **Case 3.5.1.** $v =$ **null**
        By rule DYN-ARRAY-LOOKUP-NULL, the configuration steps to $\langle H, \textbf{Error} \rangle$.
    **Case 3.5.2.** $v = \iota_\sigma$
        We do a case analysis on whether $i$ is within the array bounds or not:
        **Case 3.5.2.1.** $i \notin \text{dom}(\sigma)$
            By rule DYN-ARRAY-LOOKUP-FAIL with (2), the configuration steps to $\langle H, \textbf{Error} \rangle$.
        **Case 3.5.2.2** $i \in \text{dom}(\sigma)$
            We do a case analysis on whether the array contains read-only elements or not:
            **Case 3.5.2.2.1.** readOnlyElems($x$)
                By rule DYN-ARRAY-LOOKUP with (2), the configuration steps to $\langle H, (S, v) \rangle$, where $v = H(\iota)[\sigma(i)]$. By Lemma 4 (The heap mirrors $\Delta$), this array exists in $H$.
            **Case 3.5.2.2.1.** $\neg$readOnlyElems($x$)
                By rule DYN-ARRAY-LOOKUP-UNIQUE with (2), the configuration steps to $\langle H', (S, v) \rangle$, where $v = H(\iota)[\sigma(i)]$ and $H' = H[\iota \mapsto (H(\iota)[\sigma(i) \mapsto \textbf{null}])]$.
                By Lemma 4 (The heap mirrors $\Delta$), this array exists in $H$.

**Case 3.6.** $e = x[i] = e'$
1. $\Delta; \Gamma \vdash x : \alpha \ [\textbf{var} \ t]$ by **wf-e** (E-ARRAY-ASSIGN).
2. $S(x) = v$, and $\Delta; \Gamma \vdash v : \alpha \ [\textbf{var} \ t]$ by Lemma 1 (The stack mirrors $\Gamma$) with (1).
3. We do a case analysis on whether $v$ is **null** or a reference.
    **Case 3.6.1.** $v =$ **null**
        By rule DYN-ARRAY-ASSIGN-NULL, the configuration steps to $\langle H, \textbf{Error} \rangle$.
    **Case 3.6.2.** $v = \iota_\sigma$
        We do a case analysis on whether $i$ is within the array bounds or not:
        **Case 3.6.2.1.** $i \notin \text{dom}(\sigma)$
            By rule DYN-ARRAY-ASSIGN-FAIL with (1), the configuration steps to $\langle H, \textbf{Error} \rangle$.
        **Case 3.6.2.2.** $i \in \text{dom}(\sigma)$
            By rule DYN-ARRAY-ASSIGN with (1), the configuration steps to $\langle H', \textbf{true} \rangle$, where $H' = H[\iota \mapsto (H(\iota)[\sigma(i) \mapsto v])]$. By Lemma 4 (The heap mirrors $\Delta$), this array exists in $H$.

**Case 3.7.** $e = \textbf{let} \ y_{\sigma_1} \uplus z_{\sigma_2} = x \ \textbf{in} \ e'$
1. $\Delta; \Gamma \vdash x : \alpha \ [mod \ t']$ by **wf-e** (E-ARRAY-SPLIT).
2. $S(x) = v$, and $\Delta; \Gamma \vdash v : \alpha \ [mod \ t']$ by Lemma 1 (The stack mirrors $\Gamma$) with (1).
3. We do a case analysis on whether $v$ is **null** or a reference.
    **Case 3.7.1.** $v =$ **null**
        By rule DYN-ARRAY-SPLIT-NULL, the configuration steps to $\langle H, \textbf{Error} \rangle$.
    **Case 3.7.2.** $v = \iota_\sigma$

By rule DYN-ARRAY-SPLIT with (2), the configuration steps to $\langle H, (S', e') \rangle$, where $S' = S, x \mapsto \mathbf{null}, y \mapsto \iota_{\sigma'_1}, z \mapsto \iota_{\sigma'_2}$, and $\sigma'_1 = \sigma \circ \sigma_1$ and $\sigma'_2 = \sigma \circ \sigma_2$.

**Case 3.8.** $e = x \uplus y$

1. $\Delta; \Gamma \vdash x : \alpha \ [mod \ t']$ and $\Delta; \Gamma \vdash y : \alpha \ [mod \ t']$ by *wf-e* (E-ARRAY-MERGE).
2. $S(x) = v_1$ and $S(y) = v_2$, both of which are have types, by Lemma 1 (The stack mirrors $\Gamma$) with (1).
3. We do a case analysis on whether any of $v_1$ or $v_2$ are **null** or if they are both references.

    **Case 3.8.1.** $v_1 = \mathbf{null} \lor v_2 = \mathbf{null}$

    By rule DYN-ARRAY-MERGE-NULL, the configuration steps to $\langle H, \mathbf{Error} \rangle$.

    **Case 3.8.2.** $v_1 = \iota_{\sigma_1}$ and $v_2 = \iota'_{\sigma_2}$

    We do a case analysis on whether $\iota$ and $\iota'$ are equal:

    **Case 3.8.2.1.** $\iota \neq \iota'$.

    By rule DYN-ARRAY-MERGE-FAIL with (2), the configuration steps to $\langle H, \mathbf{Error} \rangle$.

    **Case 3.8.2.2.** $\iota = \iota'$.

    By rule DYN-ARRAY-MERGE with (2), the configuration steps to $\langle H, (S', \iota_{\sigma_1 \uplus \sigma_2}) \rangle$, where $S' = S[x \mapsto \mathbf{null}][y \mapsto \mathbf{null}]$.

**Case 3.9.** $e = \mathbf{new\ unique}\ [mod\ t](n)$

We do a case analysis on $t$:

**Case 3.9.1.** $t = \mathbf{bool}$

By rule DYN-ARRAY-NEW, the configuration steps to $\langle H, \iota \mapsto [\overline{\mathbf{false}}^n], (S, \iota_{id}) \rangle$, where $\iota$ is a fresh memory location.

**Case 3.9.2.** $t \neq \mathbf{bool}$ (meaning it must be an array)

By rule DYN-ARRAY-NEW, the configuration steps to $\langle H, \iota \mapsto [\overline{\mathbf{null}}^n], (S, \iota_{id}) \rangle$, where $\iota$ is a fresh memory location.

**Case 3.10.** $e = \mathbf{finish}\{\ \mathbf{async}\{e_1\}\mathbf{async}\{e_2\}\}; e_3$

By rule DYN-SPAWN, the configuration steps to $\langle H, (S_1, e_2) \parallel (S_2, e_2) \triangleright S, e_3 \rangle$, where $S_i$ contains all the variables from $S$ that appear in $e_i$.

**Case 3.11.** $e = \mathbf{borrow}\ x\ \mathbf{as}\ y\ \mathbf{in}\ e'$

1. $S(x) = v$ by Lemma 1 (The stack mirrors $\Gamma$) with *wf-e*.
2. By rule DYN-BORROW with (1), the configuration steps to $\langle H, (S', \mathbf{B}(e)) \rangle$, where $S' = S \bullet x \mapsto null, y \mapsto v$.

**Case 3.12.** $e = \mathbf{B}(v)$

By rule DYN-BORROW-DONE, the configuration steps to $\langle H, (S', \mathbf{true}) \rangle$, where $S'$ is $S$ without the mappings after the rightmost $\bullet$. We know there is such a $\bullet$ marker in $S$ since we have a **B** wrapper in the expression (cf. appendix B.1).

**Case 3.13.** $e = E[e']$, where $e' \neq v$

This case collects all the inductive cases, where a subexpression can be evaluated before the full expression is evaluated.

We have the following induction hypothesis (note that $e'$ is well-formed by *wf-e*, and that we have already assumed that $e'$ is not a value):

$\exists cfg'. \langle H, (S, e') \rangle \rightarrow cfg'$

We proceed by case analysis on the shape of *cfg'*:

**Case 3.13.1.** $cfg' = \langle H, \textbf{Error} \rangle$
By rule DYN-CONTEXT-ERROR, the configuration steps to $\langle H, \textbf{Error} \rangle$.

**Case 3.13.2.** $cfg' = \langle H', (S', e'') \rangle$
By rule DYN-CONTEXT, the configuration steps to $\langle H', (S', E[e'']) \rangle$.

**Case 3.13.3.** $cfg' = \langle H', A_1 \parallel A_2 \triangleright (S, e'') \rangle$
By rule DYN-SPAWN-CTX, the configuration steps to $\langle H', A_1 \parallel A_2 \triangleright (S, E[e'']) \rangle$.

## B.3 Preservation

The progress formulation and proof is mostly standard.

**Theorem 2** (Preservation).
*If $\Delta; \Gamma \vdash \langle H, A \rangle$ and $\langle H, A \rangle \rightarrow \langle H', A' \rangle$, then $\exists \Delta', \Gamma'$ such that $\Delta'; \Gamma' \vdash \langle H', A' \rangle$, and $\Delta \subseteq \Delta'$.*

Note that there are no requirements on the relation between $\Gamma$ and $\Gamma'$. Instead we rely on the reasoning in appendix B.1 to show that $\Gamma'$ only "forgets" variables that go out of scope (see Case 3.13).

Assumptions:

**wf-cfg:** $\Delta; \Gamma \vdash \langle H, A \rangle : t$
**step:** $\langle H, A \rangle \rightarrow \langle H', A' \rangle$

We start by induction over the shape of *A*.

**Case 1.** $cfg = \langle H, \textbf{Error} \rangle$
Preservation holds vacuously, since **Error** does not step.

**Case 2.** $cfg = \langle H, A_1 \parallel A_2 \triangleright (S, e) \rangle$
Induction hypotheses (note that $A_1$ and $A_2$ are well-formed by **wf-cfg**):
**IH1:** If $\langle H, A_1 \rangle \rightarrow \langle H', A_1' \rangle$, then $\exists \Delta', \Gamma'$ such that $\Delta'; \Gamma' \vdash \langle H', A_1' \rangle$, where $\Delta \subseteq \Delta'$
**IH2:** If $\langle H, A_2 \rangle \rightarrow \langle H', A_2' \rangle$, then $\exists \Delta', \Gamma'$ such that $\Delta'; \Gamma' \vdash \langle H', A_2' \rangle$, where $\Delta \subseteq \Delta'$
We proceed by case analysis over *step*. There are five rules which step the configuration.

**Case 2.1.** Rule DYN-SCHED-L
$\langle H, A_1 \parallel A_2 \triangleright (S, e) \rangle \rightarrow \langle H', A_1' \parallel A_2 \triangleright (S, e) \rangle$ since $\langle H, A_1 \rangle \rightarrow \langle H', A_1' \rangle$. By **IH1**, $\exists \Delta_1, \Gamma_1$ such that $\Delta_1; \Gamma_1 \vdash \langle H', A_1' \rangle$, and $\Delta \subseteq \Delta_1$.
We pick $\Delta'$ and $\Gamma'$ and show the premises of rule WF-CFG.
1. Pick $\Delta' = \Delta_1$ and $\Gamma = \Gamma$.
   a. $\Delta \subseteq \Delta_1$ by the induction hypothesis.
2. Show $\Delta_1 \vdash H'$
   Holds by inversion of the induction hypothesis.
3. Show $\Delta_1; \Gamma \vdash A_1' \parallel A_2 \triangleright (S, e)$
   We show the premises of rule WF-ASYNC.
   a. Show $\exists \Gamma_1. \Delta_1; \Gamma_1 \vdash A_1'$
   Holds by inversion of the induction hypothesis

  b. Show $\exists \Gamma_2.\Delta_1; \Gamma_2 \vdash A_2$
  Holds by Lemma 2 ($\Delta$ weakening) with **wf-cfg** and ɪa.
  c. Show $\Delta_1; \Gamma \vdash S$
  Holds by Lemma 2 ($\Delta$ weakening) with **wf-cfg** and ɪa.
  d. Show $\Delta_1; \Gamma \vdash e : t$
  Holds by Lemma 2 ($\Delta$ weakening) with **wf-cfg** and ɪa.

**Case 2.2.** Rule DYN-SCHED-R
Analogous to Case 2.1.

**Case 2.3.** Rule DYN-SCHED-L-FAIL
$\langle H, A_1 \parallel A_2 \triangleright (S,e) \rangle \rightarrow \langle H, \textbf{Error} \rangle$
$\Delta \vdash H$ by **wf-cfg**. $\Delta; \Gamma \vdash \textbf{Error}$ by rule WF-ERROR.

**Case 2.4.** Rule DYN-SCHED-R-FAIL
Analogous to Case 2.3.

**Case 2.5.** Rule DYN-FINISH
$\langle H, A_1 \parallel A_2 \triangleright (S,e) \rangle \rightarrow \langle H, (S,e) \rangle$
Configuration is well-formed by **wf-cfg**.

**Case 3.** $cfg = \langle H, (S,e) \rangle$
Assumption:

**wf-e**: $\Gamma \vdash e : t$, by **wf-cfg**

We proceed by induction over the shape of $e$. Each case is structured according to the following schema:

**Case 3.n.** $e = e'$
By **wf-cfg**, we have $\Delta'; \Gamma' \vdash e' : t$, which may give new assumptions.
For each applicable dynamic rule, we have $\langle H, (S, e') \rangle \rightarrow \langle H', A' \rangle$, and we need to show $\exists \Delta', \Gamma'$ such that $\Delta'; \Gamma' \vdash \langle H', A' \rangle : t$.
We prove well-formedness of the resulting configuration by proving well-formedness for each component:
1. Pick $\Gamma'$ and $\Delta'$, showing $\Delta \subseteq \Delta'$
2. Show $\Delta' \vdash H$
3. Show $\Delta'; \Gamma' \vdash A'$

In most cases, $A'$ will be a single thread $(S', e'')$, in which case step 3 above is replaced by
3. Show $\Delta'; \Gamma' \vdash S'$
4. Show $\Delta'; \Gamma' \vdash e'' : t$

Together, by rule WF-CFG, these steps imply $\Delta'; \Gamma' \vdash \langle H', A' \rangle : t$.

**Case 3.1.** $e = v$
Configuration does not step. Preservation holds vacuously.

**Case 3.2.** $e = x$
There are two applicable rules, depending on if the type of $x$ is read-only or not (and thus if we need to do a destructive read or not):

**Case 3.2.1.** Rule DYN-VAR-LOOKUP
$\langle H, (S,x) \rangle \rightarrow \langle H, (S,v) \rangle$, where $v = S(x)$.
1. Let $\Delta' = \Delta$ and $\Gamma' = \Gamma$

   (a) $\Delta \subseteq \Delta$ trivially
2. Show $\Delta \vdash H$
   Holds by *wf-cfg*.
3. Show $\Delta; \Gamma \vdash S$
   Holds by *wf-cfg*.
4. Show $\Delta; \Gamma \vdash v : t$
   Holds by Lemma 1 (The stack mirrors $\Gamma$) with *wf-cfg* and *wf-e*.

**Case 3.2.2.** Rule DYN-VAR-LOOKUP-DEST
$\langle H, (S, x) \rangle \rightarrow \langle H, (S', v) \rangle$, where $v = S(x)$ and $S' = S[x \mapsto \textbf{null}]$.
1. Let $\Delta' = \Delta$ and $\Gamma' = \Gamma$
   (a) $\Delta \subseteq \Delta$ trivially
2. Show $\Delta \vdash H$
   Holds by *wf-cfg*.
3. Show $\Delta; \Gamma \vdash S[x \mapsto \textbf{null}]$
   a. $\Delta; \Gamma \vdash v : t$ by Lemma 1 (The stack mirrors $\Gamma$) with *wf-cfg* and *wf-e*.
   b. $\Gamma \vdash \textbf{null} : t$ by rule WF-NULL.
   c. $\Delta; \Gamma \vdash S[x \mapsto \textbf{null}]$ by Lemma 3 (Stack Update) with *wf-cfg*, *wf-e*, 3a and 3b.
4. Show $\Delta; \Gamma \vdash v : t$
   Holds by Lemma 1 (The stack mirrors $\Gamma$) with *wf-cfg* and *wf-e*.

**Case 3.3.** $e = \textbf{let } x = v \textbf{ in } e'$
By *wf-e* (WF-LET), we get the following assumptions:
**A1:** $\Delta; \Gamma \vdash v : t'$.
**A2:** $\Delta; \Gamma, x : t' \vdash e' : t$
The only applicable rule is DYN-LET.
$\langle H, (S, \textbf{let } x = v \textbf{ in } e') \rangle \rightarrow \langle H, (S', e') \rangle$, where $S' = S, x \mapsto v$.
1. Let $\Delta' = \Delta$ and $\Gamma' = \Gamma, x : t'$
   (a) $\Delta \subseteq \Delta$ trivially
2. Show $\Delta \vdash H$
   Holds by *wf-cfg*.
3. Show $\Delta; \Gamma, x : t' \vdash S, x \mapsto v$
   Holds by rule WF-S-VAR with **A1** and *wf-cfg*.
4. Show $\Delta; \Gamma, x : t' \vdash e' : t$
   Holds by **A2**.

**Case 3.4.** $e = f(v)$
By *wf-e* (E-CALL), we get the following assumptions:
**A1:** $\Delta; \Gamma \vdash v : t'$, and
**A2:** $P(f) = \textbf{fun } f(x : t') : t \; e'$.
By rule WF-FUNCTION, we also get that:
**A3:** $\Delta; \Gamma, x : t' \vdash e' : t$.
The only applicable rule is DYN-CALL.
$\langle H, (S, f(v)) \rangle \rightarrow \langle H, (S', e') \rangle$, where $S' = S, x \mapsto v$.
1. Let $\Delta' = \Delta$ and $\Gamma' = \Gamma, x : t'$

   (a) $\Delta \subseteq \Delta$ trivially
2. Show $\Delta \vdash H$
   Holds by *wf-cfg*.
3. Show $\Delta; \Gamma, x : t' \vdash S, x \mapsto v$
   Holds by rule wf-s-var with **A1**.
4. Show $\Delta; \Gamma, x : t' \vdash e' : t$
   Holds by **A3**.

**Case 3.5.** $e = x \uplus y$

There are three applicable rules, depending on if $x$ and $y$ refer to the same array or not, and if either of them are **null**.

**Case 3.5.1.** Rule dyn-array-merge

$\langle H, (S, x \uplus y) \rangle \rightarrow \langle H, (S', \iota_{\sigma_1 \uplus \sigma_2}) \rangle$, where $S' = S[x \mapsto \mathbf{null}][y \mapsto \mathbf{null}]$, and $S(x) = \iota_{\sigma_1}$ and $S(y) = \iota_{\sigma_2}$.

By *wf-e* (e-array-merge), we get the following assumptions:

**A1:** $\Delta; \Gamma \vdash x : t$,

**A2:** $\Delta; \Gamma \vdash y : t$, and

1. Let $\Delta' = \Delta$ and $\Gamma' = \Gamma$
   (a) $\Delta \subseteq \Delta$ trivially
2. Show $\Delta \vdash H$
   Holds by *wf-cfg*.
3. Show $\Delta; \Gamma \vdash S[x \mapsto \mathbf{null}][y \mapsto \mathbf{null}]$
   Holds by Lemma 3 (Stack Update) with rule e-null and *wf-cfg*.
4. Show $\Delta; \Gamma \vdash \iota_{\sigma_1 \uplus \sigma_2} : t$ Holds by:
   (a) $\Delta; \Gamma \vdash \iota_{\sigma_1} : t$ by Lemma 1 (The stack mirrors $\Gamma$) with **A1** and *wf-cfg*.
   (b) $\Delta(\iota) = t$ by 4a and rule e-array.
   (c) $\Delta; \Gamma \vdash \iota_{\sigma_1 \uplus \sigma_2} : t$ by rule e-array with 4b.

**Case 3.5.2.** Rule dyn-array-merge-fail

$\langle H, (S, x \uplus y) \rangle \rightarrow \langle H, \mathbf{Error} \rangle$.

$\Delta \vdash H$ by *wf-cfg*. $\Delta; \Gamma \vdash \mathbf{Error}$ by rule wf-error.

**Case 3.5.3.** Rule dyn-array-merge-null

$\langle H, (S, x \uplus y) \rangle \rightarrow \langle H, \mathbf{Error} \rangle$.

$\Delta \vdash H$ by *wf-cfg*. $\Delta; \Gamma \vdash \mathbf{Error}$ by rule wf-error.

**Case 3.6.** $e = x[i]$

There are four applicable rules, depending on if $x$ is **null** or not, if the index is within bounds or not, and if the array referred to by $x$ contains read-only capabilities (and thus can be read non-destructively) or not.

**Case 3.6.1.** Rule dyn-array-lookup-null

$\langle H, (S, x[i]) \rangle \rightarrow \langle H, \mathbf{Error} \rangle$.

$\Delta \vdash H$ by *wf-cfg*. $\Delta; \Gamma \vdash \mathbf{Error}$ by rule wf-error.

**Case 3.6.2.** Rule dyn-array-lookup-fail

Analogous to Case 3.6.1.

**Case 3.6.3.** Rule dyn-array-lookup

$\langle H, (S, x[i]) \rangle \rightarrow \langle H, (S, v) \rangle$, where $S(x) = \iota_\sigma$ and $v = H(\iota)[\sigma(i)]$

1. Let $\Delta' = \Delta$ and $\Gamma' = \Gamma$
   (a) $\Delta \subseteq \Delta$ trivially
2. Show $\Delta \vdash H$
   Holds by **wf-cfg**.
3. Show $\Delta; \Gamma \vdash S$
   Holds by **wf-cfg**.
4. Show $\Delta; \Gamma \vdash H(\iota)[\sigma(i)] : t$
   (a) $\Delta; \Gamma \vdash \iota_\sigma : \alpha \ [mod\ t]$ by Lemma 1 (The stack mirrors $\Gamma$) with **wf-cfg**.
   (b) $\Delta; \Gamma \vdash H(\iota)[\sigma(i)] : t$ by Lemma 4 (The heap mirrors $\Delta$) with 4a and **wf-cfg**.

**Case 3.6.4.** Rule DYN-ARRAY-LOOKUP-UNIQUE
$\langle H, (S, x[i]) \rangle \rightarrow \langle H', (S, v) \rangle$, where $S(x) = \iota_\sigma$, $v = H(\iota)[\sigma(i)]$, and $H' = H[\iota \mapsto (H(\iota)[\sigma(i) \mapsto \mathbf{null}])]$.
1. Let $\Delta' = \Delta$ and $\Gamma' = \Gamma$
   (a) $\Delta \subseteq \Delta$ trivially
2. Show $\Delta \vdash H[\iota \mapsto (H(\iota)[\sigma(i) \mapsto \mathbf{null}])]$
   (a) $\Delta; \Gamma \vdash H$ by **wf-cfg**.
   (b) $\Delta; \Gamma \vdash \iota_\sigma : \alpha \ [mod\ t]$ by Lemma 1 (The stack mirrors $\Gamma$) with **wf-cfg**.
   (c) $\Delta(\iota) = \alpha \ [mod\ t]$ by inversion of 2b (WF-ARRAY).
   (d) $\Delta; \Gamma \vdash H[\iota \mapsto (H(\iota)[\sigma(i) \mapsto \mathbf{null}])]$ by Lemma 5 (Heap Array Update) with 2a, 2c and rule WF-NULL.
3. Show $\Delta; \Gamma \vdash S$
   Holds by **wf-cfg**.
4. Show $\Delta; \Gamma \vdash H(\iota)[\sigma(i)] : t$
   Analogous to Case 3.6.3.4

**Case 3.7.** $e = x[i] = v$
There are three applicable rules, depending on if $x$ is **null** or not, and if $i$ is within the bounds of the array or not.

**Case 3.7.1.** Rule DYN-ARRAY-ASSIGN-NULL
$\langle H, (S, x[i]) \rangle \rightarrow \langle H, \mathbf{Error} \rangle$.
$\Delta \vdash H$ by **wf-cfg**. $\Delta; \Gamma \vdash \mathbf{Error}$ by rule WF-ERROR.

**Case 3.7.2.** Rule DYN-ARRAY-ASSIGN-FAIL
Analogous to Case 3.7.1.

**Case 3.7.3.** Rule DYN-ARRAY-ASSIGN
$\langle H, (S, x[i] = v) \rangle \rightarrow \langle H', (S, \mathbf{true}) \rangle$, where $H' = H[\iota \mapsto (H(\iota)[\sigma(i) \mapsto v])]$.
By **wf-e** (E-ARRAY-ASSIGN), we get the following assumptions

**A1:** $\Delta; \Gamma \vdash v : t$, and

**A2:** $\Delta; \Gamma \vdash x : \alpha \ [\mathbf{var}\ t]$, and

By rule DYN-ARRAY-ASSIGN we also get:

**A3:** $S(x) = \iota_\sigma$,

1. Let $\Delta' = \Delta$ and $\Gamma' = \Gamma$
   (a) $\Delta \subseteq \Delta$ trivially
2. Show $\Delta \vdash H'$

   Holds by
   (a) $\Delta; \Gamma \vdash H$ by *wf-cfg*.
   (b) $\Delta; \Gamma \vdash \iota_\sigma : \alpha \ [\mathbf{var}\ t]$ by Lemma 1 (The stack mirrors $\Gamma$) with **A2**, **A3** and *wf-cfg*.
   (c) $\Delta(\iota) = \alpha\ [\mathbf{var}\ t]$ by inversion of 2b
   (d) $\Delta; \Gamma \vdash H[\iota \mapsto (H(\iota)[\sigma(i) \mapsto v]$ by Lemma 5 (Heap Array Update) with 2a, 2c and **A1**.
3. Show $\Delta; \Gamma \vdash S$
   Holds by *wf-cfg*.
4. Show $\Delta; \Gamma \vdash \mathbf{true} : \mathbf{bool}$
   Holds by rule WF-V-BOOL.

**Case 3.8.** $e = \mathbf{let}\ y_{\sigma_1} \uplus z_{\sigma_2} = x\ \mathbf{in}\ e'$
   There are two applicable rules, depending on if $x$ is **null** or not.

   **Case 3.8.1.** Rule DYN-ARRAY-SPLIT-NULL
   $\langle H, (S, x[i]) \rangle \rightarrow \langle H, \mathbf{Error} \rangle$.
   $\Delta \vdash H$ by *wf-cfg*. $\Delta; \Gamma \vdash \mathbf{Error}$ by rule WF-ERROR.

   **Case 3.8.2.** Rule DYN-ARRAY-SPLIT
   $\langle H, (S, \mathbf{let}\ y_{\sigma_1} \uplus z_{\sigma_2} = x\ \mathbf{in}\ e') \rangle \rightarrow \langle H, (S', e') \rangle$, where $S' = S[x \mapsto \mathbf{null}], y \mapsto \iota_{\sigma'_1}, z \mapsto \iota_{\sigma'_2}$, and $\sigma'_1 = \sigma \circ \sigma_1$, $\sigma'_2 = \sigma \circ \sigma_2$, and $S(x) = \iota_\sigma$.
   By *wf-e* (E-ARRAY-SPLIT), we get the following assumption:
   **A1:** $\Delta; \Gamma \vdash x : \alpha\ [mod\ t']$.
   **A2:** $\Delta; \Gamma, y : \alpha\ [mod\ t'], z : \alpha\ [mod\ t'] \vdash e : t$.
   1. Let $\Delta' = \Delta$ and $\Gamma' = \Gamma, y : \alpha\ [mod\ t'], z : \alpha\ [mod\ t']$
      (a) $\Delta \subseteq \Delta$ trivially
   2. Show $\Delta \vdash H$
      Holds by *wf-cfg*.
   3. Show $\Delta; \Gamma, y : \alpha\ [mod\ t'], z : \alpha\ [mod\ t'] \vdash S, x \mapsto \mathbf{null}, y \mapsto \iota_{\sigma'_1}, z \mapsto \iota_{\sigma'_2}$.
      Holds by
      (a) $\Delta; \Gamma \vdash S[x \mapsto \mathbf{null}]$ by Lemma 3 (Stack Update) with rule E-NULL and *wf-cfg*,
      (b) $\Delta \vdash \iota_\sigma : \alpha\ [mod\ t']$ by Lemma 1 (The stack mirrors $\Gamma$)
      (c) $\Delta(\iota) = \alpha\ [mod\ t']$ by inversion of 3b
      (d) $\Delta; \Gamma, y : \alpha\ [mod\ t'] \vdash \iota_{\sigma'_2} : \alpha\ [mod\ t']$, and $\Delta; \Gamma \vdash \iota_{\sigma'_1} : \alpha\ [mod\ t']$ by 3c
      (e) $\Delta; \Gamma, y : \alpha\ [mod\ t'], z : \alpha\ [mod\ t'] \vdash S[x \mapsto \mathbf{null}], y \mapsto \iota_{\sigma'_1}, z \mapsto \iota_{\sigma'_2}$ by rule WF-S-VAR (twice) with 2d and 2a.
   4. Show $\Delta; \Gamma, y : \alpha\ [mod\ t'], z : \alpha\ [mod\ t'] \vdash e' : t$
      Holds by **A2**

**Case 3.9.** $e = \mathbf{borrow}\ x\ \mathbf{as}\ y\ \mathbf{in}\ e'$
   There only applicable rule is DYN-BORROW.
   $\langle H, (S, \mathbf{borrow}\ x\ \mathbf{as}\ y\ \mathbf{in}\ e') \rangle \rightarrow \langle H, (S', \mathbf{B}(e')) \rangle$, where $S' = S \bullet x \mapsto \mathbf{null}, y \mapsto v$, and $v = S(x)$
   By *wf-e* (E-BORROW), we get that:
   **A1:** $\Delta; \Gamma \vdash x : \alpha\ [mod\ t']$

**A2:** $\Delta; \Gamma \bullet x : \textbf{buried}[mod\ t'], y : t'' \vdash e' : t$, where $t'' = \textbf{borrowed}\ [mod\ t'] \vee t'' = \textbf{borrowed}\ [\textbf{val}\ R(t')]$

Note that we do not care about the shape of $t''$ in this case.

1. Let $\Delta' = \Delta$ and $\Gamma' = \Gamma \bullet x : \textbf{buried}[mod\ t'], y : t''$
   (a) $\Delta \subseteq \Delta$ trivially
2. Show $\Delta \vdash H$
   Holds by *wf-cfg*.
3. Show $\Delta; \Gamma \bullet x : \textbf{buried}\ [mod\ t'], y : t'' \vdash S \bullet x \mapsto \textbf{null}, y \mapsto v$
   Holds by
   (a) $\Delta; \Gamma \vdash S$ by *wf-cfg*,
   (b) $\Delta; \Gamma \vdash v : \alpha\ [mod\ t']$ by Lemma 1 (S mirrors $\Gamma$) with **A1** and 3a.
   (c) $\Delta; \Gamma \bullet x : \textbf{buried}\ [mod\ t'] \vdash S \bullet x \mapsto \textbf{null}$, by rule WF-S-BORROW with 3a and A1.
   (d) $\Delta; \Gamma \bullet x : \textbf{buried}\ [mod\ t'], y : t'' \vdash S \bullet x \mapsto \textbf{null}, y \mapsto v$, by rule WF-S-VAR with 3c and 3b.
4. Show $\Delta; \Gamma \bullet x : \textbf{buried}\ [mod\ t'], y : t'' \vdash e' : t$
   Holds by rule E-BORROWED with **A2**.

**Case 3.10.** $e = \textbf{new unique}\ [mod\ t'](n)$

There only applicable rule is DYN-ARRAY-NEW.

$\langle H, (S, \textbf{new unique}\ [mod\ t'](n)) \rangle \rightarrow \langle H', (S, \iota_{id}) \rangle$, where $\iota$ is fresh and $H' = H, \iota \mapsto [v_0, ..., v_{n-1}]$. $v_i$ is either **false** or **null**, depending on the shape of $t'$.

1. Let $\Delta' = \Delta, \iota : \textbf{unique}\ [mod\ t]$ and $\Gamma' = \Gamma$
   (a) $\Delta \subseteq \Delta'$ by construction.
2. Show $\Delta, \iota : \textbf{unique}\ [mod\ t] \vdash H, \iota \mapsto [v_0, ..., v_{n-1}]$
   We do a case analysis on whether $t'$ is **bool** or an array:
   (a) $\Delta' \vdash H$ by Lemma 2 ($\Delta$ weakening) with *wf-cfg*.
   (b) **Case :** $t' = \textbf{bool}$:
      The values in the new array are all **false**.
      $\overline{\Delta'; \Gamma \vdash \textbf{false} : \textbf{bool}}^n$ by rule E-BOOL.
   (c) **Case :** array($t'$):
      The values in the new array are all **null**.
      $\overline{\Delta'; \Gamma \vdash \textbf{null} : \textbf{bool}}^n$ by rule E-NULL.
   (d) $\Delta' \vdash H, \iota \mapsto [v_0, ..., v_{n-1}]$ by rule WF-H-ADD with 2a, and 2b or 2c. Note also that dom($\Delta'$) = dom($H$).
3. Show $\Delta'; \Gamma \vdash S$
   Holds by Lemma 2 ($\Delta$ weakening) with *wf-cfg*.
4. Show $\Delta'; \Gamma \vdash \iota_{id} : \textbf{unique}\ [mod\ t']$
   Holds by:
   (a) $\Delta'(\iota) = \textbf{unique}\ [mod\ t']$ trivially.
   (b) $\Delta'; \Gamma \vdash \iota_{id} : unique\ [mod\ t']$ by rule E-ARRAY with 4a.

**Case 3.11.** $e = \textbf{finish}\{\textbf{async}\{e_1\}\textbf{async}\{e_2\}\}; e_3$

There only applicable rule is DYN-SPAWN.

$\langle H, (S, \textbf{finish}\{\ \textbf{async}\{\ e_1\}\textbf{async}\{\ e_2\}\}; e_3) \rangle \rightarrow \langle H, (S_1, e_1)\ \|\ (S_2, e_2) \triangleright (S, e_3) \rangle$, where $S_1 = [x \mapsto v \mid x \in \textsf{fv}(e_1) \wedge S(x) = v]$ and $S_2 = [x \mapsto v \mid x \in \textsf{fv}(e_2) \wedge S(x) = v]$.

By *wf-e* (E-FINISH-ASYNC), we get the following assumptions:

**A1:** $\Delta; \Gamma \vdash e_1 : t_1$.

**A2:** $\Delta; \Gamma \vdash e_2 : t_2$.

**A3:** $\Delta; \Gamma \vdash e_3 : t$.

1. Let $\Delta' = \Delta$ and $\Gamma' = \Gamma$.
   (a) $\Delta \subseteq \Delta'$ trivially.
2. Show $\Delta \vdash H$
   Holds by *wf-cfg*.
3. Show $\Delta; \Gamma \vdash (S_1, e_1) \parallel (S_2, e_2) \triangleright (S, e_3)$
   By rule WF-ASYNC we need to pick $\Gamma_1$ and $\Gamma_2$ so that the two newly spawned threads are well-formed under them.
   (a) Mirroring the structure of their respective stack, we pick $\Gamma_1 = [x : t \mid x \in \text{dom}(S_1) \wedge \Gamma(x) = t]$ and $\Gamma_2 = [x : t \mid x \in \text{dom}(S_2) \wedge \Gamma(x) = t]$, preserving the order of the variables. Note that there are no $\bullet$ markers, as these are only used for buried variables, which cannot appear in the free variables of an expression.
   (b) Clearly, $\Delta; \Gamma_1 \vdash S_1$ and $\Delta; \Gamma_2 \vdash S_2$, since the environments were constructed using the stacks as models.
   (c) Since $\Gamma_1$ and $\Gamma_2$ only contain variables from $e_1$ and $e_2$ respectively, and by **A1** and **A2**, we have that $\Delta; \Gamma_1 \vdash e_1 : t_1$ and $\Delta; \Gamma_2 \vdash e_2 : t_2$. Together with 3b we get $\Delta; \Gamma_1 \vdash (S_1, e_1) : t_1$ and $\Delta; \Gamma_2 \vdash (S_2, e_2) : t_2$.
   (d) $\Delta; \Gamma \vdash S$ by *wf-cfg*
   (e) $\Delta; \Gamma \vdash (S_1, e_1) \parallel (S_2, e_2) \triangleright (S, e_3) : t$ by rule WF-ASYNC with 3c, 3d and **A3**.

**Case 3.12.** $e = \mathbf{B}(v)$

There only applicable rule is DYN-BORROW-DONE.

$\langle H, (S, \mathbf{B}(v)) \rangle \rightarrow \langle H, (S', v) \rangle$, where $S'$ is $S$ without the mappings after the rightmost $\bullet$.

By *wf-e* (E-BORROWED), we get the following assumption:

**A1:** $\Delta; \Gamma \vdash v : t$.

1. Let $\Delta' = \Delta$ and $\Gamma'$ as $\Gamma$ without the mappings after the rightmost $\bullet$.
   (a) $\Delta \subseteq \Delta'$ trivially.
2. Show $\Delta \vdash H$.
   Holds by *wf-cfg*.
3. Show $\Delta; \Gamma' \vdash S'$
   Holds since $\Delta; \Gamma \vdash S$ (by *wf-cfg*) and $\Gamma'$ is created from $\Gamma$ via the same transformation as $S'$ is created from $S$.
4. Show $\Delta; \Gamma' \vdash v : t$
   Holds, since $\Delta; \Gamma \vdash v : t$ by **A1**, and no values depend on the contents of $\Gamma$.

**Case 3.13.** $e = E[e']$, where $e' \neq v$

This case collects all the inductive cases, where a subexpression can be evaluated before the full expression is evaluated.

We have the following induction hypothesis (note that $e'$ is well-formed by *wf-e*, and that we have already assumed that $e'$ is not a value):

$\langle H, (S, e') \rangle \rightarrow cfg' \implies \exists \Delta', \Gamma'. \Delta'; \Gamma' \vdash cfg' : t$, where $\Delta \subseteq \Delta'$

There are three applicable rules.

**Case 3.13.1.** Rule DYN-CONTEXT-ERROR
$\langle H, (S, E[e']) \rangle \to \langle H, \mathbf{Error} \rangle$
$\Delta \vdash H$ by **wf-cfg**. $\Delta; \Gamma \vdash \mathbf{Error}$ by rule WF-ERROR.

**Case 3.13.2.** Rule DYN-CONTEXT
$\langle H, (S, E[e']) \rangle \to \langle H', (S', E[e'']) \rangle$
By the induction hypothesis, we know that $\Delta'; \Gamma' \vdash \langle H', (S', e'') \rangle : t'$, where $\Delta; \Gamma \vdash e' : t'$.

1. Let $\Delta' = \Delta'$ and $\Gamma' = \Gamma'$
   (a) $\Delta \subseteq \Delta'$ by the induction hypothesis.
2. Show $\Delta' \vdash H'$.
   Holds by the induction hypothesis.
3. Show $\Delta'; \Gamma' \vdash S'$
   Holds by the induction hypothesis.
4. Show $\Delta'; \Gamma' \vdash E[e''] : t$
   (a) $\Delta'; \Gamma \vdash E[e'] : t$ by Lemma 2 ($\Delta$ weakening) with **wf-cfg**.
   (b) $\Delta'; \Gamma' \vdash e'' : t'$ (where $\Delta; \Gamma \vdash e' : t'$), by the induction hypothesis.
   (c) In order to go from $\Delta'; \Gamma \vdash E[e'] : t$ to $\Delta'; \Gamma' \vdash E[e''] : t$ we replace $e'$ with $e''$ (which under $\Gamma'$ has the same type as $e'$) and replace $\Gamma$ with $\Gamma'$. In doing so, we need to make sure that the subexpressions in $E[e'']$ which are not in $e''$ are still well-formed under this new $\Gamma'$. The previous cases of the proof show three possible shapes of $\Gamma'$:
   i. $\Gamma' = \Gamma$: The subexpressions of $E[e'']$ that are not in $e''$ well-formed by **wf-e**.
   ii. $\Gamma \subseteq \Gamma'$: The subexpressions of $E[e'']$ that are not in $e''$ are well-formed by a traditional weakening argument.
   iii. $\Gamma'$ is $\Gamma$ without the mappings after the rightmost •, and the evaluation of $e'$ removed the innermost **B** wrapper.
   By the reasoning in appendix B.1, assume that there are $d+1$ • markers in $\Gamma$, and that the **B** wrapper removed from $e'$ wrapped an expression at depth $d + 1$. This means that $E[e']$, and therefore $E[e'']$, is at a borrowing depth which is at most $d$. With $\Gamma'$, we disallow variables at a borrowing depth greater than $d$. But the subexpressions that are not in $e''$ have not yet been evaluated, meaning they cannot contain **B** wrappers, so any free variables must also appear at borrowing depth $d$. Thus, the subexpressions that are not in $e''$ are well-formed under $\Gamma'$ as well.
   (d) $\Delta'; \Gamma' \vdash E[e''] : t$ by 3a, 3b and 3c.

**Case 3.13.3.** $cfg' = \langle H', A_1 \parallel A_2 \triangleright (S, e'') \rangle$
By rule DYN-SPAWN-CTX, the configuration steps to $\langle H', A_1 \parallel A_2 \triangleright (S, E[e'']) \rangle$.

## B.4 Disjointness of Mutable Arrays

In order to prove that no two array capabilities can update the same array element, we define a function caps which extracts a multiset of all the capabilities in a running system. It traverses the activity structure and visits all the running threads.

In the case of a fork, we collect the capabilities of the two running activities, and subtract them from the capabilities of the waiting thread (remember that we are building a multiset, so the subtraction is not cancelled out by the union). In other words, we don't collect capabilities from the waiting thread if they are already being used by the running threads; we consider this kind of aliasing benign. In the case of a single thread, we traverse the stack and current expression of that thread.

$$\text{caps}(\langle H, A \rangle) = \text{caps}_A(H; A)$$

$$\text{caps}_A(H; \textbf{Error}) = \emptyset$$
$$\text{caps}_A(H; A_1 \parallel A_2 \triangleright (S, e)) = C \cup (\text{caps}_A(H; (S, e)) \setminus C)$$
$$\text{where } C = \text{caps}_A(H; A_1) \cup \text{caps}_A(H; A_2)$$
$$\text{caps}_A(H; (S, e)) = \text{caps}_S(H; \emptyset; S) \cup \text{caps}_e(H; e)$$

When traversing a stack, we keep a set $B$ of variables that have been buried so that we can omit these from the collected capabilities; these cannot be used until borrowing stops. We collect non-buried capabilities when we find them.

$$\text{caps}_S(H; B; \epsilon) = \emptyset$$
$$\text{caps}_S(H; B; S \bullet x_t \mapsto \textbf{null}) = \text{caps}_S(H; B \cup \{x\}, S)$$
$$\text{caps}_S(H; B; S, x_t \mapsto \iota_\sigma) = \text{caps}_S(H; B; S) \cup \text{caps}_e(H; \iota_\sigma^t) \quad x \notin B$$
$$\text{caps}_S(H; B; S, x_t \mapsto v) = \text{caps}_S(H; B; S) \quad\quad\quad x \in B$$
$$\text{caps}_S(H; B; S, x_t \mapsto v) = \text{caps}_S(H; B; S) \quad\quad\quad v \neq \iota_\sigma$$

Finally, when traversing an expression, we are interested in finding subexpressions that are capabilities. When we find such a capability, we record its location $\iota$, its translation function $\sigma$, and its type, and also record all the capabilities recursively accessible in the array guarded by this capability ($v^t$ takes $\iota_\sigma$ to $\iota_\sigma^t$ and leaves booleans untouched).

$$\text{caps}_e(H; \iota_\sigma^{\alpha \, [mod \, t]}) = (\iota, \sigma, \alpha \, [mod \, t]) \cup$$
$$\bigcup_{i \in \text{dom}(\sigma)} \text{caps}_e(H; v^t), \text{where } v = H(\iota)[\sigma(i)])\}$$
$$\text{caps}_e(H; E[e]) = \text{caps}_e(H; e)$$
$$\text{caps}_e(H; \_) = \emptyset$$

We use $c.\iota$, $c.\sigma$, and $c.t$ respectively to extract the array identifier, the index translation function, and the type of an array capability $c$. The array disjointness property states that any two capabilities in the system either refer to different arrays, refer to different subparts of the same array, or are both read-only:

$$\text{arrayDisjointness}(cfg) \equiv$$
$$\forall c_1, c_2 \in \text{caps}(cfg) \, .$$
$$c_1.\iota \neq c_2.\iota \, \vee \, \text{rng}(c_1.\sigma) \cap \text{rng}(c_2.\sigma) = \emptyset \, \vee \, \text{read}(c_1.t) \wedge \text{read}(c_2.t)$$

**Reference Capabilities for Safe Parallel Array Programming**

**Theorem 3** (Preservation of Array Disjointness).
*If $\Delta; \Gamma \vdash \textit{cfg}$, and* arrayDisjointness(*cfg*)*, and cfg $\rightarrow$ cfg′, then* arrayDisjointness(*cfg′*)

At any time during the execution of a well-formed *Arr-O-Matic* program, we have a set of reachable capabilities caps(*cfg*) for which the arrayDisjointness property holds. An important observation in our proof is that if *cfg* steps to *cfg'*, and the new set of capabilities caps(*cfg′*) is equal to or a subset of caps(*cfg*), arrayDisjointness holds for this new configuration. For example, when a reference is moved by a destructive read from a variable on the stack to a value, the set of capabilities does not change. We will refer to this observation in the proofs as the *subset property*.

Assumptions:

**wf-cfg**  $\Delta; \Gamma \vdash \textit{cfg} : t$

**step:** $\langle H, A \rangle \rightarrow \langle H', A' \rangle$

We start by induction over the shape of *A*.

**Case 1.** $\textit{cfg} = \langle H, \mathbf{Error} \rangle$

The arrayDisjointness property holds vacuously, since **Error** does not step.

**Case 2.** $\textit{cfg} = \langle H, A_1 \parallel A_2 \triangleright (S, e) \rangle$

Induction hypotheses (note that $A_1$ and $A_2$ are well-formed by **wf-cfg** and that *cfg* fulfils the arrayDisjointness property):

**IH1:** If $\langle H, A_1 \rangle \rightarrow \langle H', A_1' \rangle$, then the arrayDisjointness property holds for $\langle H', A_1' \rangle$

**IH2:** If $\langle H, A_2 \rangle \rightarrow \langle H', A_2' \rangle$, then arrayDisjointness property holds for $\langle H', A_2' \rangle$

We proceed by case analysis over **step**. There are five rules which step the configuration.

**Case 2.1.** Rule DYN-SCHED-L

$\langle H, A_1 \parallel A_2 \triangleright (S, e) \rangle \rightarrow \langle H', A_1' \parallel A_2 \triangleright (S, e) \rangle$ since $\langle H, A_1 \rangle \rightarrow \langle H', A_1' \rangle$. **IH1** gives us that arrayDisjointness($\langle H', A_1' \rangle$). If caps($\langle H', A_1' \rangle$) $\subseteq$ caps($\langle H, A_1 \rangle$) then arrayDisjointness holds by the subset property. We observe that caps($\langle H', A_1' \rangle$) can only grow by the creation of a new array (which is benign), or by duplicating, splitting or merging capabilities already in caps($\langle H, A_1 \rangle$). If a duplicated, split or merged capability overlaps with a capability *c* from $A_2$ or *S*, the source of this capability would also have overlapped with *c*, contradicting the assumption that arrayDisjointness holds for $\langle H, A_1 \parallel A_2 \triangleright (S, e) \rangle$. Thus, arrayDisjointness holds for $\langle H', A_1' \parallel A_2 \triangleright (S, e) \rangle$.

**Case 2.2.** Rule DYN-SCHED-R

Analogous to Case 2.1.

**Case 2.3.** Rule DYN-SCHED-L-FAIL

$\langle H, A_1 \parallel A_2 \triangleright (S, e) \rangle \rightarrow \langle H, \mathbf{Error} \rangle$

Error states trivially fulfil the arrayDisjointness property.

**Case 2.4.** Rule DYN-SCHED-R-FAIL

Analogous to Case 2.3.

**Case 2.5.** Rule DYN-FINISH

$\langle H, A_1 \parallel A_2 \triangleright (S, e) \rangle \rightarrow \langle H, (S, e) \rangle$

arrayDisjointness holds by the subset property, as $\text{caps}(\langle H,(S,e)\rangle) \subseteq \text{caps}(\langle H, A_1 \parallel A_2 \triangleright (S,e)\rangle)$.

**Case 3.** $\textit{cfg} = \langle H,(S,e)\rangle$

New assumption:

***wf-e*:** $\Gamma \vdash e : t$, by ***wf-cfg***

We proceed by induction over the shape of $e$. To avoid cluttering the presentation, we prove all of the inductive cases (which are analogous) together in a single case. This means that in the rest of the cases, we assume that all subexpressions that can be selected using the evaluation context $E$ are values.

**Case 3.1.** $e = v$

Configuration does not step. Preservation holds vacuously.

**Case 3.2.** $e = x$

There are two applicable rules, depending on if the type of $x$ is read-only or not (and thus if we need to do a destructive read or not):

**Case 3.2.1.** Rule DYN-VAR-LOOKUP

$\langle H,(S,x)\rangle \rightarrow \langle H,(S,v)\rangle$

By rule DYN-VAR-LOOKUP, the type of $x$ is read-only. If the introduction of $v$ into the set of capabilities breaks arrayDisjointness, so would the existence of $x$ in $S$ in the original configuration, which would contradict our assumptions.

**Case 3.2.2.** Rule DYN-VAR-LOOKUP-DEST

$\langle H,(S,x)\rangle \rightarrow \langle H,(S[x \mapsto \textbf{null}],v)\rangle$

arrayDisjointness holds by the subset property, since $x$ is destructively read from the stack to the expression.

**Case 3.3.** $e = \textbf{let}\ x = v\ \textbf{in}\ e'$

The only applicable rule is DYN-LET.

$\langle H,(S, \textbf{let}\ x = v\ \textbf{in}\ e')\rangle \rightarrow \langle H,(S',e)\rangle$, where $S' = S, x \mapsto v$

arrayDisjointness holds by the subset property, since $v$ is moved from the expression to the stack.

**Case 3.4.** $e = f(v)$

The only applicable rule is DYN-CALL.

$\langle H,(S,f(v))\rangle \rightarrow \langle H,(S',e)\rangle$, where $S' = S, x \mapsto v$

arrayDisjointness holds by the subset property, as $v$ is moved from the expression to the stack.

**Case 3.5.** $e = x \uplus y$

There are three applicable rules, depending on if $x$ and $y$ refer to the same array or not, and if either of them are **null**.

**Case 3.5.1.** Rule DYN-ARRAY-MERGE

$\langle H,(S, x \uplus y)\rangle \rightarrow \langle H,(S', \iota_{\sigma_1 \uplus \sigma_2})\rangle$, where $S' = S[x \mapsto \textbf{null}][y \mapsto \textbf{null}]$, and $S(x) = \iota_{\sigma_1}$ and $S(y) = \iota_{\sigma_2}$

The new capability $\iota_{\sigma_1 \uplus \sigma_2}$ has the same access rights as $x$ and $y$, which are nullified in the process. If $\iota_{\sigma_1 \uplus \sigma_2}$ overlaps with some other capability, so would $x$ or $y$ in the original configuration, which would contradict our assumptions.

**Case 3.5.2.** Rule DYN-ARRAY-MERGE-FAIL
$\langle H, (S, x \uplus y) \rangle \rightarrow \langle H, \textbf{Error} \rangle$
Error states trivially fulfil the arrayDisjointness property.

**Case 3.5.2.** Rule DYN-ARRAY-MERGE-NULL
$\langle H, (S, x \uplus y) \rangle \rightarrow \langle H, \textbf{Error} \rangle$
Error states trivially fulfil the arrayDisjointness property.

**Case 3.6.** $e = x[i]$
There are four applicable rules, depending on if $x$ is **null** or not, if the index is within bounds or not, and if the array referred to by $x$ contains read-only capabilities (and thus can be read non-destructively) or not.

**Case 3.6.1.** Rule DYN-ARRAY-LOOKUP-NULL
$\langle H, (S, x[i]) \rangle \rightarrow \langle H, \textbf{Error} \rangle$
Error states trivially fulfil the arrayDisjointness property.

**Case 3.6.2.** Rule DYN-ARRAY-LOOKUP-FAIL
Analogous to Case 3.6.1.

**Case 3.6.3.** Rule DYN-ARRAY-LOOKUP
$\langle H, (S, x[i]) \rangle \rightarrow \langle H, (S, v) \rangle$, where $S(x) = \iota_\sigma$ and $v = H(\iota)[\sigma(i)]$
By rule DYN-ARRAY-LOOKUP, the type of $x[i]$ is read-only. If the introduction of $v$ into the set of capabilities breaks arrayDisjointness, so would the existence of $x[i]$ in $H$ in the original configuration, which would contradict our assumptions.

**Case 3.6.4.** Rule DYN-ARRAY-LOOKUP-UNIQUE
$\langle H, (S, x[i]) \rangle \rightarrow \langle H', (S, v) \rangle$, where $S(x) = \iota_\sigma$, $v = H(\iota)[\sigma(i)]$, and $H' = H[\iota \mapsto (H(\iota)[\sigma(i) \mapsto \textbf{null}])]$.
arrayDisjointness holds by the subset property, since $x[i]$ is destructively read from the heap to the expression.

**Case 3.7.** $e = x[i] = v$
There are three applicable rules, depending on if $x$ is **null** or not, and if $i$ is within the bounds of the array or not.

**Case 3.7.1.** Rule DYN-ARRAY-ASSIGN-NULL
$\langle H, (S, x[i] = v) \rangle \rightarrow \langle H, \textbf{Error} \rangle$
Error states trivially fulfil the arrayDisjointness property.

**Case 3.7.2.** Rule DYN-ARRAY-ASSIGN-FAIL
Analogous to Case 3.7.1.

**Case 3.7.3.** Rule DYN-ARRAY-ASSIGN
$\langle H, (S, x[i] = v) \rangle \rightarrow \langle H', (S, \textbf{true}) \rangle$, where $H' = H[\iota \mapsto (H(\iota)[\sigma(i) \mapsto v])]$
arrayDisjointness holds by the subset property, since $v$ is moved from the expression to the heap.

**Case 3.8.** $e = \textbf{let } y_{\sigma_1} \uplus z_{\sigma_2} = x \textbf{ in } e'$
There are two applicable rules, depending on if $x$ is **null** or not.

**Case 3.8.1.** Rule DYN-ARRAY-SPLIT-NULL
$\langle H, (S, x[i]) \rangle \rightarrow \langle H, \textbf{Error} \rangle$.
Error states trivially fulfil the arrayDisjointness property.

**Case 3.8.2.** Rule DYN-ARRAY-SPLIT

$\langle H, (S, \text{let } y_{\sigma_1} \uplus z_{\sigma_2} = x \text{ in } e')\rangle \to \langle H, (S', e')\rangle$, where $S' = S[x \mapsto \text{null}], y \mapsto \iota_{\sigma_1}, z \mapsto \iota_{\sigma_2}$, and $\sigma'_1 = \sigma \circ \sigma_1, \sigma'_2 = \sigma \circ \sigma_2$, and $S(x) = \iota_\sigma$.

We drop one capability ($x$) and introduce two new ones ($y$ and $z$). Since we compose the translation functions $\sigma_1$ and $\sigma_2$ with $\sigma$, $y$ and $z$ will not have access to anything that $x$ did not. By assumption, $x$ does not overlap with any other capabilities in the configuration, so neither will $y$ and $z$. By rule E-ARRAY-SPLIT, the ranges of the translation functions $\sigma_1$ and $\sigma_2$ are disjoint, meaning $y$ and $z$ will have access to disjoint parts of $\iota$.

Thus, the introduction of $y$ and $z$ does not violate the arrayDisjointness property.

**Case 3.9.** $e = \text{borrow } x \text{ as } y \text{ in } e'$

There only applicable rule is DYN-BORROW.

$\langle H, (S, \text{borrow } x \text{ as } y \text{ in } e')\rangle \to \langle H, (S', \mathbf{B}(e'))\rangle$, where $S' = S \bullet x \mapsto \text{null}, y \mapsto v$ and $v = S(x)$.

We move the capability of $x$ into $y$, but at the same time bury the original reference (by $\bullet\ x \mapsto \text{null}$). Burying $x$ means that it won't show up in $\text{caps}_S(S', \emptyset)$. If $x$ was a read-only capability, there could be other read-only aliases left, but by *wf-e* rule E-BORROW, the modifier of $y$ can only be **var** if the modifier of $x$ was **var**, so we cannot introduce a mutable alias of an immutable capability. Thus, the only difference between $\text{caps}(\langle H, (S, \text{borrow } x \text{ as } y \text{ in } e')\rangle)$ and $\text{caps}(\langle H, (S', \mathbf{B}(e'))\rangle)$ is that the type of the moved capability is now **borrowed**, and possibly read-only. This does not affect the arrayDisjointness property negatively.

**Case 3.10.** $e = \text{new unique } [mod\ t](n)$

There only applicable rule is DYN-ARRAY-NEW.

$\langle H, (S, \text{new unique } [mod\ t](n))\rangle \to \langle H', (S, \iota_\sigma)\rangle$, where $H' = H, \iota \mapsto [v_0, ..., v_{n-1}]$.

We create a fresh capability referring to a new array, which does not refer to any other arrays. No other capabilities are affected, so the arrayDisjointness property still holds.

**Case 3.11.** $e = \text{finish}\{\text{async}\{e_1\}\text{async}\{e_2\}\}; e_3$

There only applicable rule is DYN-SPAWN.

$\langle H, (S, \textit{finish}\{\textit{async}\{e_1\}\textit{async}\{e_2\}\}; e_3)\rangle \to \langle H, (S_1, e_1)\ ||\ (S_2, e_2) \triangleright (S, e_3)\rangle$, where $S_1 = [x \mapsto v | x \in \text{fv}(e1) \land S(x) = v]$ and $S_2 = [x \mapsto v | x \in \text{fv}(e2) \land S(x) = v]$

arrayDisjointness holds by the subset property, since $S_1$ and $S_2$ are both "substacks" of $S$, and the definition of $\text{caps}_A()$ for forks removes capabilities in $S$ if they appear in $S_1$ or $S_2$.

**Case 3.12.** $e = \mathbf{B}(v)$

There only applicable rule is DYN-BORROW-DONE.

$\langle H, (S, \mathbf{B}(v))\rangle \to \langle H, (S', \mathbf{true})\rangle$, where $S'$ is $S$ without the mappings after the rightmost $\bullet$.

$S'$ is strictly smaller than $S$, but will recover the variable that was buried when the **B** was introduced. By the reasoning in appendix B.1, all the aliases of the previously buried variable are dropped from the stack, and there can be no aliases on the heap. Since we also drop the expression $v$, there are no aliases in the expression of the resulting configuration either. Because of this, there can be no aliases of the previously buried variable that did not also exist before the variable was buried, meaning arrayDisjointness holds for the resulting configuration.

**Case 3.13.** $e = E[e']$, where $e' \neq v$

This case collects all the inductive cases, where a subexpression can be evaluated before the full expression is evaluated.

We have the following induction hypothesis (note that $e'$ is well-formed by **wf-e**, and that we have already assumed that $e'$ is not a value):

If $\langle H, (S, e) \rangle \rightarrow cfg'$ then arrayDisjointness holds for $cfg'$.

We proceed by case analysis on the shape of $cfg'$

**Case 3.13.1.** $cfg' = \langle H, \textbf{Error} \rangle$

By rule DYN-CONTEXT-ERROR, the configuration steps to $\langle H, \textbf{Error} \rangle$.

Error states trivially fulfil the arrayDisjointness property.

**Case 3.13.2.** $cfg' = \langle H', (S', e'') \rangle$

By rule DYN-CONTEXT, the configuration steps to $\langle H', (S', E[e'']) \rangle$.

By the induction hypothesis, caps($\langle H', (S', e'') \rangle$), contain no overlapping capabilities.

The only difference between this set of capabilities and caps($\langle H', (S', E[e'']) \rangle$) is the capabilities in the expression. Looking at the definition of $E$, we see that the only time there is more than one subexpression is in the **let** case, namely the body of the **let**. Since we don't start evaluating the body until we have calculated the value of the new variable, there can be no free capabilities in this subexpression. Thus, caps($\langle H', (S', e'') \rangle$) = caps($\langle H', (S', E[e'']) \rangle$), and arrayDisjointness holds by the subset property.

**Case 3.13.3.** $cfg' = \langle H', A_1 \parallel A_2 \triangleright (S, e'') \rangle$

By rule DYN-SPAWN-CTX, the configuration steps to $\langle H', A_1 \parallel A_2 \triangleright (S, E[e'']) \rangle$.

Analogous to Case 3.13.2.

## B.5 Lemmas

**Lemma 1 (The stack mirrors $\Gamma$)**:

$\Delta; \Gamma \vdash S \wedge \Delta; \Gamma \vdash x : t \implies S(x) = v \wedge \Delta; \Gamma \vdash v : t$

*Explanation:* If the stack is well-formed under $\Gamma$ and $\Gamma$ maps $x$ to type $t$, then the stack maps $x$ to a value $v$ of type $t$.

*Proof:* By rule WF-S-VAR we se that the stack has the same domain as $\Gamma$, and that each variable is mapped to a value of the same type as in $\Gamma$. Note also that $\Delta; \Gamma \vdash x : t$ implies that $x$ is not buried, so rule WF-S-BORROW does not apply.

**Lemma 2 ($\Delta$ weakening)**:

$\Delta \subseteq \Delta' \wedge \Delta; \Gamma \vdash \langle H, A \rangle : t \implies \Delta' \vdash H \wedge \Delta'; \Gamma \vdash A : t$

*Explanation:* If a configuration $cfg$ is well-formed under $\Delta$ and there is a $\Delta'$ where $\Delta$ is fully subsumed by $\Delta'$, then the constituents of the configuration are also well-formed under $\Delta'$. Note that the whole configuration is not well-formed under $\Delta'$, as rule WF-CFG also requires dom($\Delta$) = dom($H$), which does not hold in general.

*Proof:* Proven by induction over the shape of the heap, the activities *A*, and over the internal stacks and expressions *S* and *e*. The only interesting cases are rules WF-H-ADD and E-ARRAY, which both require $\Delta(\iota) = t$, for some respective *t*. For a $\Delta'$, where $\Delta \subseteq \Delta'$, if $\Delta(\iota) = t$, then $\Delta'(\iota) = t$.

*Corollary:* By inversion of $\Delta'; \Gamma \vdash A : t$, we get that each subpart of an activity (that is, any internal *S* and *e*) is also well-formed under weakening. We will refer to this Lemma when proving these properties as well.

**Lemma 3 (Stack Update)** :

$\Delta; \Gamma \vdash S \wedge S(x) = v \wedge \Delta; \Gamma \vdash v : t \wedge \Delta; \Gamma \vdash v' : t \Rightarrow \Delta; \Gamma \vdash S[x \mapsto v']$

*Explanation:* If the stack *S* is well-formed under $\Delta$ and $\Gamma$, the stack *S* maps the name *x* to the value *v*, *v* has type *t* under $\Delta$ and $\Gamma$, and $v'$ has type *t* under $\Delta$ and $\Gamma$, then the stack *S* will be well-formed under $\Delta$ and $\Gamma$ when the name *x* has been updated to map to the value $v'$.

*Proof:* By rule WF-S-VAR we get that a stack $S, x \mapsto v$ is well-formed under $\Delta; \Gamma, x : t$ given $\Delta \vdash v : t$ and $\Delta; \Gamma \vdash S$. This gives that also $S, x \mapsto v'$ is well-formed under $\Delta; \Gamma, x : t$, since $\Delta \vdash v' : t$.

**Lemma 4 (The heap mirrors $\Delta$)**

$\Delta \vdash H \wedge \Delta; \Gamma \vdash \iota_\sigma : \alpha \, [mod \, t] \wedge \text{dom}(\Delta) = \text{dom}(H) \implies H(\iota)[\sigma(i)] = v \wedge \Delta; \Gamma \vdash v : t$

*Explanation:* If the heap is well-formed, the array $\iota_\sigma$ has type $\alpha \, [mod \, t]$ and the domain of $\Delta$ and the domain of *H* are equal, the result of a lookup in the array will be a value *v* of type *t*.

*Proof:* Given $\Delta \vdash H$, $\Delta(\iota) = \alpha \, [mod \, t]$ and rule WF-H-ADD we know that the heap stores a number of elements of type *t* in a sequence at the address $\iota$. By that we also know that the lookup $H(\iota)[\sigma(i)]$ will return a value *v* with the type *t*. Since the domains of $\Delta$ and *H* are the same, there cannot be an $\iota$ in $\Delta$ that is not also an address on the heap.

**Lemma 5 (Heap Array Update)**

$\Delta; \Gamma \vdash H \wedge \Delta; \Gamma \vdash \iota_\sigma : \alpha[\mathbf{var} \, t] \wedge \Delta; \Gamma \vdash v : t \wedge \text{dom}(\Delta) = \text{dom}(H) \implies \Delta; \Gamma \vdash H[\iota \mapsto (H(\iota)[\sigma(i) \mapsto v])]$

*Explanation:* If the heap is well-formed, the array $\iota_\sigma$ has type $\alpha \, [\mathbf{var} \, t]$ and *v* has type *t*, then the heap update $H[\iota \mapsto (H(\iota)[\sigma(i) \mapsto v])]$ will be well-formed.

*Proof:* By rule WF-H-ADD we get that an array at location $\iota$ in a heap is well-formed under $\Delta$ given that $\Delta(\iota) = \alpha \, [mod \, t]$ and that the elements all have type *t* under $\Delta$. This gives that also $H[\iota \mapsto (H(\iota)[\sigma(i) \mapsto v])]$ is still well-formed under $\Delta$, since $\Delta \vdash v : t$. By the same reasoning as in Lemma 4, we know that $\iota$ is indeed a location in the heap, since the domains of $\Delta$ and *H* are the same.

Reference Capabilities for Safe Parallel Array Programming

**About the authors**

**Beatrice Åkerblom** is a PhD student at Stockholm University, in Stockholm, Sweden. Contact her at: beatrice@dsv.su.se.

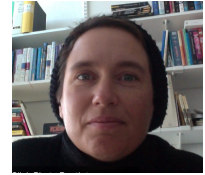

**Elias Castegren** is a postdoctoral research associate at KTH Royal Institute of Technology, in Stockholm, Sweden. Contact him at: eliasca@kth.se.

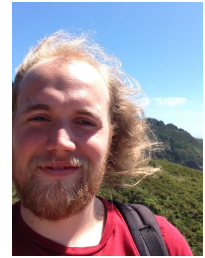

**Tobias Wrigstad** is an Associate Professor at Uppsala University, in Uppsala, Sweden. Contact him at: tobias.wrigstad@it.uu.se.

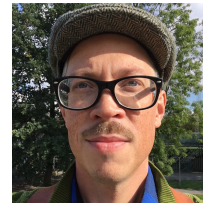